\documentclass[%
reprint,
superscriptaddress,
showpacs,
amsmath,amssymb,
prc,
floatfix ]%
{revtex4-1}

\usepackage{color}
\usepackage{CJK}

\usepackage{graphicx}
\usepackage{dcolumn}
\usepackage{bm}
\usepackage{hyperref}

\allowdisplaybreaks

\begin{document}

\begin{CJK*}{UTF8}{}
\title{Microscopic self-consistent description of induced fission dynamics: finite temperature effects}
\CJKfamily{gbsn}
\author{Jie Zhao (赵杰)}%
\affiliation{Microsystem and Terahertz Research Center and Insititute of Electronic Engineering, 
	China Academy of Engineering Physics, Chengdu 610200, Sichuan, China}
\author{Tamara Nik\v{s}i\'c}%
\affiliation{Physics Department, Faculty of Science, University of Zagreb, Bijeni\v{c}ka Cesta 32,
              Zagreb 10000, Croatia}             
\author{Dario Vretenar}%
\affiliation{Physics Department, Faculty of Science, University of Zagreb, Bijeni\v{c}ka Cesta 32,
              Zagreb 10000, Croatia}
\CJKfamily{gbsn}              
\author{Shan-Gui Zhou (周善贵)}%
 \affiliation{CAS Key Laboratory of Theoretical Physics, Institute of Theoretical Physics, Chinese Academy of Sciences, Beijing 100190, China}
 \affiliation{School of Physical Sciences, University of Chinese Academy of Sciences, Beijing 100049, China}
 \affiliation{Center of Theoretical Nuclear Physics, National Laboratory of Heavy Ion Accelerator, Lanzhou 730000, China}
 \affiliation{Synergetic Innovation Center for Quantum Effects and Application, Hunan Normal University, Changsha 410081, China}

\date{\today}

\begin{abstract}
The dynamics of induced fission of $^{226}$Th is investigated in a theoretical framework based on the finite-temperature 
time-dependent generator coordinate method (TDGCM) in the Gaussian overlap approximation (GOA).
The thermodynamical collective potential and inertia tensor at temperatures in the interval 
$T=0 - 1.25$ MeV are calculated using the self-consistent multidimensionally constrained relativistic mean field (MDC-RMF) model, 
based on the energy density functional DD-PC1. 
Pairing correlations are treated in the BCS approximation with a separable pairing force of finite range.
Constrained RMF+BCS calculations are carried out in the collective space of axially symmetric quadrupole 
and octupole deformations for the asymmetric fissioning nucleus $^{226}$Th. The collective Hamiltonian is determined by the 
temperature-dependent free energy surface and perturbative cranking inertia tensor, and the TDGCM+GOA is used to 
propagate the initial collective state in time.
The resulting charge and mass fragment distributions are analyzed as functions of the internal excitation energy. The model can qualitatively 
reproduce the empirical triple-humped structure of the fission charge and mass distributions already at $T=0$, but 
the precise experimental position of the asymmetric peaks 
and the symmetric-fission yield can only be accurately reproduced when the potential and inertia tensor of the collective Hamiltonian
are determined at finite temperature, in this particular case between $T=0.75$ MeV and $T=1$ MeV.
\end{abstract}

\maketitle

\end{CJK*}

\bigskip

\section{Introduction~\label{sec:Introduction}}

Distributions of fission fragments present basic fission observables that can be used to asses and validate theoretical methods \cite{Schunck2016_RPP79-116301}. 
For instance, the experimental study of seventy short-lived radioactive isotopes in the region $85 \leq Z \leq 92$ 
has shown that the charge and mass yields are symmetric in the lighter mass region, 
whereas the yields tend to be asymmetric for heavier nuclei and relatively low 
excitation energies \cite{Schmidt2000_NPA665-221}.
The charge and mass distributions remain asymmetric up to Cf \cite{Andreyev2013_RMP85-1541}.
The probability of symmetric fission increases with excitation energy because of the weakening 
of shell effects \cite{Isaev2008_NPA809-1,Ryzhov2011_PRC83-054603,Sen2017_PRC96-064609,
Leguillon2016_PLB761-125,Demetriou2001_NPA695-95,Hirose2017_PRL119-222501,Simutkin2014_NDS119-331,
Ramos2018_PRC97-054612,Naik2018_PRC97-014614}. 

A microscopic theoretical approach capable of predicting fission fragment distributions starting from the initial state of the 
compound nucleus is the 
time-dependent generator coordinate method (TDGCM)~\cite{Schunck2016_RPP79-116301,Berger1991_CPC63-365}. 
In the Gaussian overlap approximation (GOA) the GCM Hill-Wheeler equation reduces to a local, time-dependent 
Schr\"odinger-like equation in the space of collective coordinates.
For a choice of collective coordinates, the essential inputs are the potential and inertia tensor that 
can be determined microscopically in a self-consistent mean-field deformation-constrained calculation.
Most applications of the TDGCM+GOA to nuclear fission dynamics have been based on non-relativistic Skyrme 
and Gogny functionals~\cite{Berger1991_CPC63-365,Goutte2005_PRC71-024316,Younes2012_LLNL-TR-586678,
Regnier2018_CPC225-180,Regnier2017_EPJWC146-04043,Regnier2016_PRC93-054611,
Regnier2016_CPC200-350,Zdeb2017_PRC95-054608}. More recently, relativistic energy density functionals \cite{Vretenar2005_PR409-101,Meng2006_PPNP57-470,Meng2016_WorldSci} have also been employed 
in the description of fission properties of heavy and superheavy 
nuclei~\cite{Zhou2016_PS91-063008,Burvenich2004_PRC69-014307,Blum1994_PLB323-262,Zhang2003_CPL20-1694,
Bender2003_RMP75-121,Lu2006_CPL23-2940,Li2010_PRC81-064321,Abusara2010_PRC82-044303,
Abusara2012_PRC85-024314,Lu2012_PRC85-011301R,Lu2014_PRC89-014323,Zhao2015_PRC91-014321,
Agbemava2017_PRC95-054324,Prassa2012_PRC86-024317}. 
Triaxial and octupole deformations~\cite{Zhao2015_PRC92-064315}, and the effect of coupling between shape 
and pairing degrees of freedom~\cite{Zhao2016_PRC93-044315} on dynamic spontaneous fission paths and half-lives 
were analyzed using the multidimensionally-constrained relativistic mean-field (MDC-RMF)~\cite{Lu2014_PRC89-014323} 
and  the relativistic Hartree Bogoliubov (MDC-RHB) model~\cite{Zhao2017_PRC95-014320}.
The first study of fission dynamics that used the TDGCM+GOA based on a relativistic energy 
density functional was recently reported in Ref.~\cite{Tao2017_PRC96-024319}, where
the effect of pairing correlations on the charge yields and total kinetic energy of fission fragments was examined.

In all applications of the TDGCM+GOA to studies of induced fission dynamics \cite{Goutte2005_PRC71-024316,Younes2012_LLNL-TR-586678,Regnier2016_PRC93-054611,Zdeb2017_PRC95-054608,Tao2017_PRC96-024319}, the 
collective potential and inertia tensor have been calculated at 
zero temperature. However, as the internal excitation energy increases, one expects that both 
the potential energy surface (PES) and the mass parameters exhibit significant modifications. 
Finite-temperature (FT) nuclear density functional theory (DFT) \cite{Goodman1981_NPA352-30} 
provides a convenient framework in which the evolution of a PES and inertia tensor with excitation 
energy can be described.
Several studies of the dependence of PESs and fission barriers on excitation energy have been carried out using the 
Finite-Temperature Hartree-Fock-Bogoliubov (FT-HFB) method based on non-relativistic
Skyrme~\cite{Zhu2016_PRC94-024329,Schunck2015_PRC91-034327,McDonnell2013_PRC87-054327,
McDonnell2014_PRC90-021302R,Pei2009_PRL102-192501} and Gogny functionals~\cite{Martin2009_IJMPE18-861}.
The effect of FT on perturbative cranking inertia tensors has also been investigated in the FT-HFB framework 
\cite{Zhu2016_PRC94-024329,Martin2009_IJMPE18-861}. 
Exploratory studies of FT effects on induced fission yield distributions using 
semi-classical approaches have been reported in Refs.~\cite{Ivanyuk2018_PRC97-054331,Randrup2013_PRC88-064606,Pasca2016_PLB760-800}.  
In this work we present the first microscopic investigation of finite temperature effects on 
induced fission dynamics using the TDGCM+GOA collective model. 
The theoretical framework and method are introduced in Sec.~\ref{sec:model}.
The details of the calculation for the illustrative example of $^{226}$Th, the results for deformation energy landscapes, 
inertia tensor, as well as the charge and mass yield distributions are described and discussed in Sec.~\ref{sec:results}.
Sec.~\ref{sec:summary} contains a summary of the principal results.

\section{\label{sec:model}The method}
Assuming that the compound nucleus is in a state of thermal equilibrium at temperature $T$, 
it can be described by the finite temperature (FT) Hartree-Fock-Bogoliubov (HFB) theory~\cite{Goodman1981_NPA352-30,Egido1986_NPA451-77}. 
In the grand-canonical ensemble, the expectation value of any operator $\hat{O}$ is given by an ensemble average 
\begin{equation}
\langle \hat{O} \rangle = \textrm{Tr} ~[ \hat{D}\hat{O} ],
\end{equation}
where $\hat{D}$ is the density operator:
\begin{equation}
\hat{D} = {1 \over Z} ~ e^{ -\beta \left( \hat{H}-\lambda \hat{N} \right) }\; .
\end{equation}
$Z$ is the grand partition function, $\beta=1/k_{B}T$ with the Boltzmann constant $k_{B}$, $\hat{H}$ is the Hamiltonian of the system, 
$\lambda$ denotes the chemical potential, and $\hat{N}$ is the particle number operator. In the present study we employ the relativistic 
mean-field (RMF) model for the particle-hole channel, while pairing correlations are treated in  
the BCS approximation. The Dirac single-nucleon equation 
\begin{equation}
\hat{h} \psi_{k}(\bm{r}) = \epsilon_{k} \psi_{k}(\bm{r}),
\end{equation}
is determined by the Hamiltonian
\begin{equation}
\label{eq:hamiltonian}
 \hat{h} = \bm{\alpha} \cdot \bm{p} + \beta[M+S(\bm{r})] + V_{0}(\bm{r}) + \Sigma_R(\bm{r}),
\end{equation}
where, for the relativistic energy-density functional DD-PC1~\cite{Niksic2008_PRC78-034318}, 
the scalar potential, vector potential, and rearrangement terms read
\begin{eqnarray}
       S & = & \alpha_{S}(\rho) \rho_{S} + \delta_{S}\triangle\rho_{S}, \nonumber \\
 V_{0} & = & \alpha_{V}(\rho) \rho_{V} + \alpha_{TV}(\rho) \vec{\rho}_{TV} \cdot \vec{\tau} 
             + e{1-\tau_3 \over 2} A_{0}, \nonumber \\
\Sigma_R & = & {1\over2} {\partial \alpha_{S} \over \partial \rho} \rho_{S}^2
	     + {1\over2} {\partial \alpha_{V} \over \partial \rho} \rho_{V}^2
	     + {1\over2} {\partial \alpha_{TV} \over \partial \rho} \rho_{TV}^2\; ,
\end{eqnarray}
respectively. $M$ is the nucleon mass, $\alpha_{S}({\rho})$, $\alpha_{V}({\rho})$,  and $\alpha_{TV}({\rho})$ are
density-dependent couplings for different space-isospace channels, $\delta_{S}$ is the coupling constant of the derivative term,
and $e$ is the electric charge. In the finite-temperature RMF+BCS approximation the single-nucleon densities 
$\rho_{S}$ (scalar-isoscalar density), $\rho_{V}$ (time-like component of the 
isoscalar current), and $\rho_{TV}$  (time-like component of the isovector current), are defined by the following relations:
\begin{equation} 
\rho_{S} = \sum_{k} \bar{\psi}_{k}(\bm{r}) \psi_{k}(\bm{r}) [v_{k}^{2} (1-f_{k}) + u_{k}^{2} f_{k}],
\end{equation}
\begin{equation}
\rho_{V} = \sum_{k} \bar{\psi}_{k}(\bm{r}) \gamma^{0} \psi_{k}(\bm{r}) [v_{k}^{2} (1-f_{k}) + u_{k}^{2} f_{k}],
\end{equation}
\begin{equation}
{\rho}_{TV} = \sum_{k} \bar{\psi}_{k}(\bm{r}) \vec{\tau} \gamma^{0} \psi_{k}(\bm{r}) [v_{k}^{2} (1-f_{k}) + u_{k}^{2} f_{k}],
\end{equation}
where $f_k$ is the thermal occupation probability of a quasiparticle state
\begin{equation}
f_{k} = {1 \over 1+e^{\beta E_k}},
\end{equation}
and $\beta=1/k_{B} T$. $E_k=[(\epsilon_{k} - \lambda)^{2} + \Delta_{k}^2]^{1/2}$ 
is the quasiparticle energy, and $\lambda$ is the Fermi level.
$v_{k}^{2}$ are the BCS occupation probabilities 
\begin{equation}
v_{k}^{2} = {1 \over 2} \left( 1 - {\epsilon_{k} - \lambda \over E_{k}} \right),
\end{equation}
and $u_{k}^{2} = 1 - v_{k}^{2}$. The gap equation at finite temperature reads
\begin{equation}
\Delta_{k} = {1 \over 2} \sum_{k^\prime >0} V^{pp}_{k\bar{k}k^\prime \bar{k^\prime}} {\Delta_{k^\prime} \over E_{k^\prime}} (1-2f_{k}^\prime). 
\end{equation}
In the particle-particle channel we use a separable pairing force of finite range \cite{Tian2009_PLB676-44}:
\begin{equation}
V(\mathbf{r}_1,\mathbf{r}_2,\mathbf{r}_1^\prime,\mathbf{r}_2^\prime) = G_0 ~\delta(\mathbf{R}-
\mathbf{R}^\prime) P (\mathbf{r}) P(\mathbf{r}^\prime) \frac{1}{2} \left(1-P^\sigma\right),
\label{pairing}
\end{equation}
where $\mathbf{R} = (\mathbf{r}_1+\mathbf{r}_2)/2$ and $\mathbf{r}=\mathbf{r}_1- \mathbf{r}_2$
denote the center-of-mass and the relative coordinates, respectively. $P(\mathbf{r})$ reads 
\begin{equation}
P(\mathbf{r})=\frac{1}{\left(4\pi a^2\right)^{3/2}} e^{-\mathbf{r}^2/4a^2}.
\end{equation}
The two parameters of the interaction were originally 
adjusted to reproduce the density dependence of the pairing gap in nuclear matter at the
Fermi surface calculated with the D1S parameterization of the Gogny force~\cite{Berger1991_CPC63-365}.

The entropy of the compound nuclear system is computed using the relation:
\begin{equation}
S = -k_{B} \sum_{k} \left[ f_{k} \ln f_{k} + (1 - f_{k}) \ln (1 - f_{k}) \right].
\end{equation}
The thermodynamical potential relevant for an analysis of finite-temperature deformation effects is the Helmholtz free energy 
$F=E(T)-TS$, evaluated at constant temperature $T$ \cite{Schunck2015_PRC91-034327}.
$E(T)$ is the binding energy of the deformed nucleus, and 
the deformation-dependent energy landscape is obtained in a self-consistent finite-temperature mean-field 
calculation with constraints on the mass multipole moments $Q_{\lambda\mu} = r^\lambda Y_{\lambda \mu}$.
The nuclear shape is parameterized by the deformation parameters
\begin{equation}
 \beta_{\lambda\mu} = {4\pi \over 3AR^\lambda} \langle Q_{\lambda\mu} \rangle.
\end{equation}
The shape is assumed to be invariant under the exchange of the $x$ and $y$ axes, 
and all deformation parameters $\beta_{\lambda\mu}$ with even $\mu$ can be included simultaneously.
The self-consistent RMF+BCS equations are solved by an expansion in the 
axially deformed harmonic oscillator (ADHO) basis~\cite{Gambhir1990_APNY198-132}.
In the present study calculations have been performed 
in an ADHO basis truncated to $N_f = 20$ oscillator shells.
For details of the MDC-RMF model we refer the reader to Ref.~\cite{Lu2014_PRC89-014323}.

In the TDGCM+GOA nuclear fission is modeled as a slow adiabatic process driven by only a few collective degrees of
freedom~\cite{Regnier2016_PRC93-054611}. The dynamics is described by a local, time-dependent Schr\"odinger-like equation 
in the space of collective coordinates $\bm{q}$,
\begin{equation}
i\hbar \frac{\partial g(\bm{q},t)}{\partial t} = \hat{H}_{\rm coll} (\bm{q}) g(\bm{q},t).
\label{eq:TDGCM}
\end{equation}
The Hamiltonian $\hat{H}_{\rm coll} (\bm{q})$ reads
\begin{equation}
\hat{H}_{\rm coll} (\bm{q}) = - {\hbar^2 \over 2} \sum_{ij} {\partial \over \partial q_i} B_{ij}(\bm{q}) {\partial \over \partial q_j} + V(\bm{q}),
\label{eq:Hcoll}
\end{equation}
where $V(\bm{q})$ and $B_{ij}(\bm{q})=\mathcal{M}^{-1}(\bm{q})$ are the collective potential and mass tensor,  
both determined by microscopic self-consistent mean-field calculations based on universal energy density functionals. 
$g(\bm{q},t)$ is the complex wave function of the collective variables $\bm{q}$. 

The collective space is divided into an inner region with a single nuclear density distribution, and an external region that contains the two fission fragments. The set of scission configurations defines the hyper-surface that separates the two regions.  
The flux of the probability current through this
hyper-surface provides a measure of the probability of observing a given pair of fragments at time $t$.
Each infinitesimal surface element is associated with a given pair of fragments $(A_L, A_H)$, where $A_L$ and $A_H$ denote the 
lighter and heavier fragment, respectively.
The integrated flux $F(\xi,t)$ for a given surface element $\xi$ is defined as \cite{Regnier2018_CPC225-180}
\begin{equation}
F(\xi,t) = \int_{t_0}^{t} \int_{\xi} \bm{J}(\bm{q},t) \cdot d\bm{S}, 
\label{eq:flux}
\end{equation}
where $\bm{J}(\bm{q},t)$ is the current
\begin{equation}
\bm{J}(\bm{q},t) = {\hbar \over 2i} \bm{B}(\bm{q}) [g^{*}(\bm{q},t) \nabla g(\bm{q},t) - g(\bm{q},t) \nabla g^{*}(\bm{q},t)].
\end{equation}
The yield for the fission fragment with mass $A$ is defined by 
\begin{equation}
Y(A) \propto \sum_{\xi \in \mathcal{A}} \lim_{t \rightarrow \infty} F(\xi,t).
\end{equation}
The set $\mathcal{A}(\xi)$ contains all elements belonging to the scission hyper-surface such that one of the fragments has mass number $A$.

The inertia tensor is calculated in the finite-temperature perturbative cranking approximation \cite{Zhu2016_PRC94-024329,Martin2009_IJMPE18-861}:
\begin{equation}
\mathcal{M}^{Cp} = \hbar^{2} M_{(1)}^{-1} M_{(3)} M_{(1)}^{-1},
\label{eq:per_mass}
\end{equation}
with 
\begin{widetext}
\begin{eqnarray}
[M_{(k)}]_{ij,T} =&& {1 \over 2} \sum_{\mu \neq \nu} 
	      \langle 0 | \hat{Q}_{i} | \mu \nu \rangle
	     \langle \mu \nu | \hat{Q}_{j} | 0 \rangle 
	     \left\{ { (u_{\mu} u_{\nu} - v_{\mu} v_{\nu})^{2} \over (E_{\mu} - E_{\nu})^{k} } 
	      \left[ \tanh\left( {E_{\mu} \over 2k_{B}T}  \right) - \tanh\left( {E_{\nu} \over 2k_{B}T} \right) \right]  \right\}  \nonumber \\
	&&+ {1 \over 2} \sum_{\mu \nu} 
	     \langle 0 | \hat{Q}_{i} | \mu \nu \rangle
	     \langle \mu \nu | \hat{Q}_{j} | 0 \rangle 
	     \left\{ { (u_{\mu} v_{\nu} + u_{\nu} v_{\mu})^{2} \over (E_{\mu} + E_{\nu})^{k} } 
	     \left[ \tanh\left( {E_{\mu} \over 2k_{B}T}  \right) + \tanh\left( {E_{\nu} \over 2k_{B}T} \right) \right]  \right\}.
\label{eq:per-mass}
\end{eqnarray}
\end{widetext}

The starting point of the dynamical calculation is the choice of the collective wave packet $g(\bm{q},t=0)$.
We build the initial state as a Gaussian superposition of the quasi-bound states $g_k$, 
\begin{equation}
g(\bm{q},t=0) = \sum_{k} \exp\left( { (E_k - \bar{E} )^{2} \over 2\sigma^{2} } \right) g_{k}(\bm{q}),
\label{eq:initial-state}
\end{equation}
where the value of the parameter $\sigma$ is set to 0.5 MeV. The collective states $\{ g_{k}(\bm{q}) \}$ 
are solutions of the stationary eigenvalue equation in which the original collective potential $V(\bm{q})$ is replaced by a 
new potential $V^{\prime} (\bm{q})$ that is obtained by extrapolating the inner potential barrier with a quadratic form (see 
Ref.~\cite{Regnier2018_CPC225-180} for details). In the following 
we denote the average energy of the collective initial state by $E_{\rm coll.}^{*}$, and its value will usually be chosen 
about 1 MeV above the highest fission barrier. The mean energy $\bar{E}$ in Eq.~(\ref{eq:initial-state}) is then adjusted iteratively in 
such a way that $\langle g(t=0)| \hat{H}_{\rm coll} | g(t=0) \rangle = E_{\rm coll.}^{*}$.
\section{\label{sec:results}Induced fission dynamics of $^{226}$Th: results and discussion}

As in our first illustrative application of the TDGCM+GOA framework to a description of induced fission dynamics 
\cite{Tao2017_PRC96-024319}, we consider the case of $^{226}$Th and 
analyze the temperature dependence of fission barriers, perturbative cranking inertia tensors, and 
distribution of charge and mass yields. In the present study the collective coordinates are the axially symmetric quadrupole 
deformation parameter $\beta_{20}$ and octupole deformation parameter $\beta_{30}$. The starting point is a large-scale 
deformation-constrained finite-temperature self-consistent RMF+BCS calculation of the potential energy surface and 
single-nucleon wave functions. In the particle-hole channel we employ the relativistic energy functional DD-PC1~\cite{Niksic2008_PRC78-034318}. 
As noted in Sec.~\ref{sec:model}, the parameters of the finite range separable pairing force were originally adjusted to 
reproduce the pairing gap at the Fermi surface in symmetric nuclear matter as calculated with the Gogny D1S force. However, 
a number of RMF-based studies have shown that in finite nuclei the strength parameters of this force need to be fine-tuned, 
especially for heavy nuclei~\cite{Wang2013_PRC87-054331,Afanasjev2013_PRC88-014320}. Here  
the strengths have been adjusted to reproduce the empirical pairing gaps of $^{226}$Th.
The assumption is that the fissioning nucleus is in thermal equilibrium at temperature $T$. The self-consistent RMF+BCS 
calculation provides a deformation energy surface $F(\bm{q})$, and variations of the free energy 
between two points $\bm{q}_1$ and $\bm{q}_2$ are given by  
$\delta F |_{T} = F(\bm{q}_{1}, T) - F(\bm{q}_{2}, T)$~\cite{Schunck2015_PRC91-034327}. The internal excitation energy 
$E_{\rm int.}^{*}$ of a nucleus at temperature $T$ is defined as the difference between the total binding energy of the equilibrium 
RMF+BCS minimum at temperature $T$ and at $T=0$. 

In a second step the computer code FELIX (version 2.0)~\cite{Regnier2018_CPC225-180} is used for the TDGCM+GOA 
time evolution. The time step is $\delta t=5\times 10^{-4}$ zs. 
The charge and mass distributions are calculated after $2\times10^{5}$ time steps, corresponding to 100 zs.
The scission configurations are defined by using the Gaussian neck operator $\hat{Q}_{N}=\exp[-(z-z_{N})^{2} / a_{N}^{2}]$, 
where $a_{N}=1$ fm and $z_{N}$ is the position of the neck~\cite{Younes2009_PRC80-054313}.
We define the pre-scission domain by $\langle \hat{Q}_{N} \rangle>2$ and consider the frontier of this domain as the scission contour.
Just as in our pervious study of Ref.~\cite{Tao2017_PRC96-024319}, the parameters of the additional imaginary absorption potential that takes into account the escape of the collective wave packet  in the domain outside the region of calculation \cite{Regnier2018_CPC225-180} are: the absorption rate $r=20\times 10^{22}$ s$^{-1}$, and the width of the absorption band $w=1.5$.
Following Ref.~\cite{Regnier2016_PRC93-054611}, the fission yields are obtained by convoluting the raw 
flux with a Gaussian function of the number of particles. The width is set to 4 units for the mass yields, and 1.6 for the charge yields.

\subsection{\label{subsec:barriers}Temperature-dependent fission barriers and interia tensors}
\begin{figure}
 \includegraphics[width=0.48\textwidth]{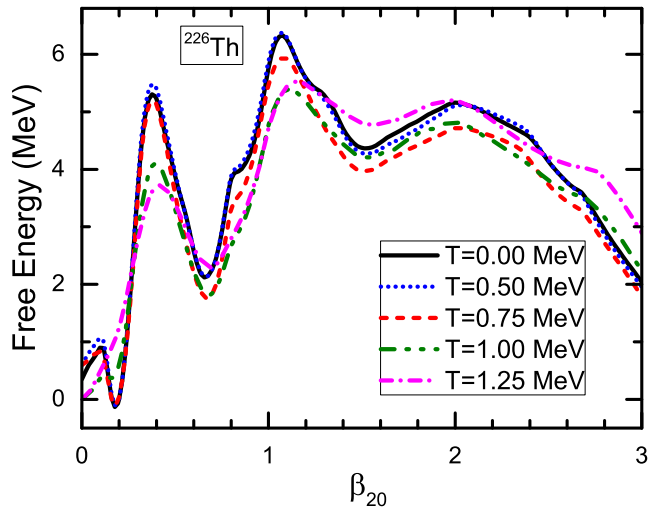}
\caption{(Color online)~\label{fig:Th226_PES_1D}%
Free energy (in MeV) along the least- energy fission pathway in $^{226}$Th for finite temperatures $T=0.0, 0.5, 0.75, 1.0, 1.25$ MeV. 
All curves are normalized to their values at equilibrium minimum.
}
\end{figure}

\begin{figure}
 \includegraphics[width=0.48\textwidth]{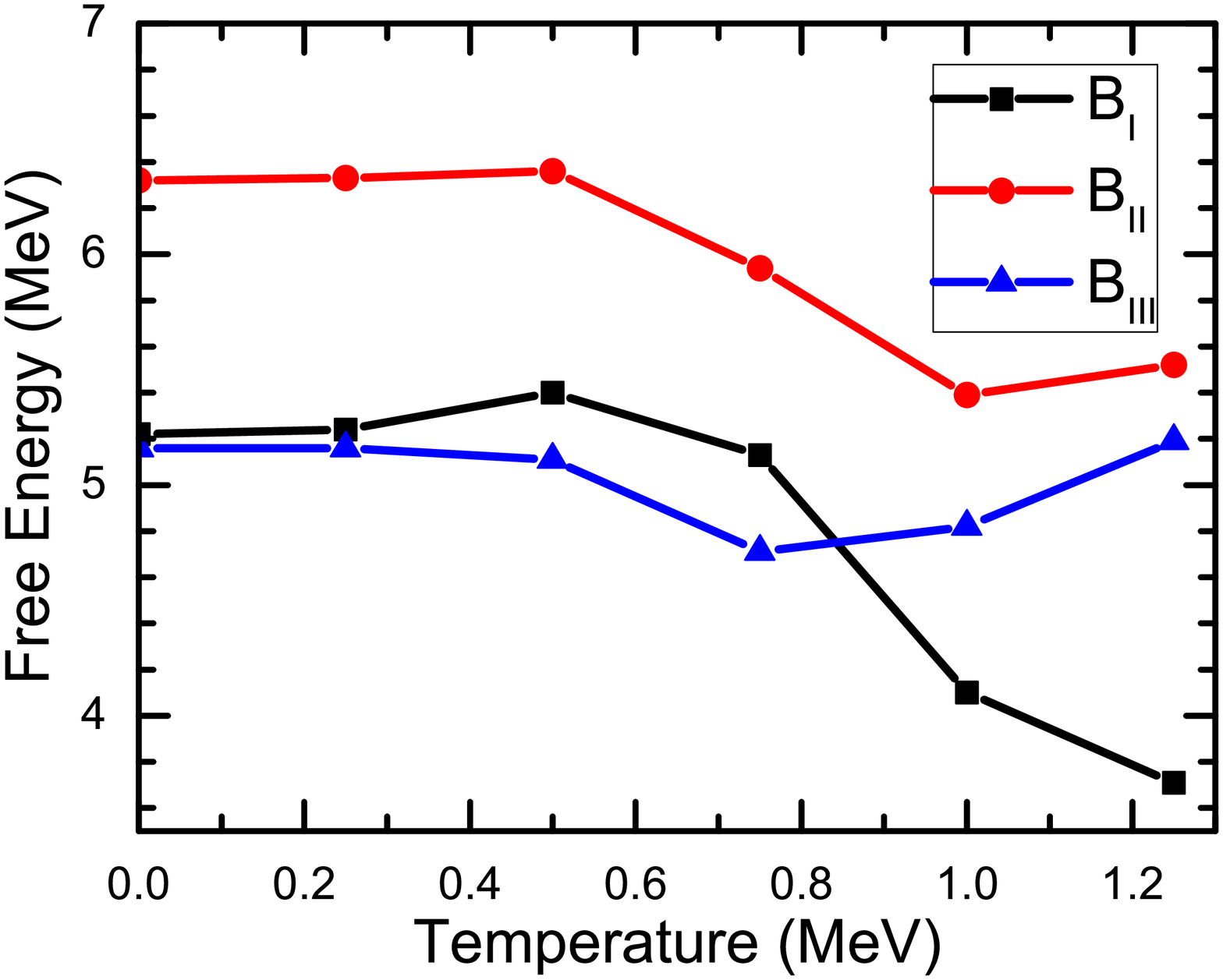}
\caption{(Color online)~\label{fig:Th226_Barriers}%
Evolution of the first ($B_{\textrm{I}}$), second ($B_{\textrm{II}}$), and third ($B_{\text{III}}$) barrier 
heights in the free energy of $^{226}$Th, as functions of temperature.
}
\end{figure}

\begin{figure}
 \includegraphics[width=0.48\textwidth]{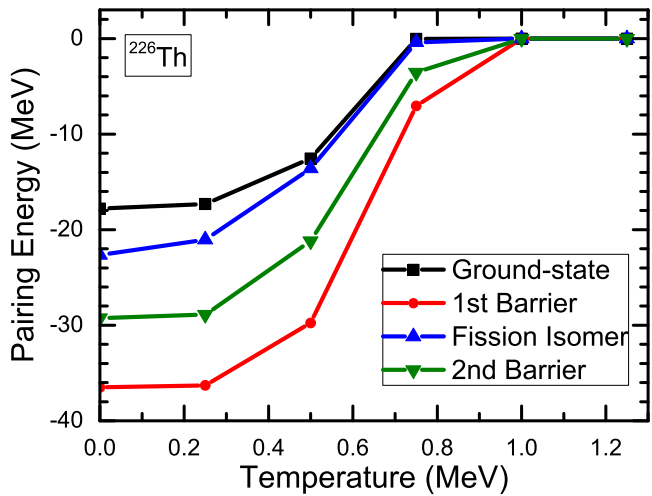}
\caption{(Color online)~\label{fig:Th226_Pairing}%
Temperature  dependence of the pairing energy in the RMF+BCS equilibrium minimum, in the fission isomer, and at the top of the first and 
second barrier in $^{226}$Th.
}
\end{figure}

Figure \ref{fig:Th226_PES_1D} displays the free energy of $^{226}$Th along the least-energy fission pathway for temperatures ranging 
between zero and 1.25 MeV.
The heights of the fission barriers as functions of temperature $T$ are plotted in Fig.~\ref{fig:Th226_Barriers}.
At $T=0$ the mean-field equilibrium state is located at $(\beta_{20},\beta_{30}) \sim (0.20,0.15)$.
Similar to the results obtained with the functional PC-PK1~\cite{Zhao2010_PRC82-054319} in Ref.~\cite{Tao2017_PRC96-024319}, 
a triple-humped barrier is predicted along the static fission path with the barrier heights $5.22$, $6.32$, and $5.16$ MeV 
from the inner to the outer barrier, respectively. One notices that the 
free energy curves do not change significantly for temperatures $T < 0.75$ MeV,
except for a modest increase of the height of the first and second barriers.
The barriers start decreasing as temperature increases beyond $T=0.75$ MeV, and at these higher temperatures 
the nucleus exhibits a spherical equilibrium shape. 
We note that although the second ($B_{II}$) and third ($B_{III}$) barriers increase slightly when $T \geq 1$ MeV,  
the depths of the second and third potential wells decrease with temperature for all $T$. At $T= 0.5, 0.75, 1.0$, and $1.25$ MeV 
the corresponding internal excitation energies $E_{\rm int.}^{*}$ 
are: $2.58$, $8.71$, $16.56$, and $27.12$ MeV, respectively.

The evolution of the barrier heights as function of temperature, shown in Fig.~\ref{fig:Th226_Barriers}, 
can be attributed to different rates of damping of pairing correlations and shell effects, as 
discussed in Ref.~\cite{Schunck2015_PRC91-034327}.
In Fig~\ref{fig:Th226_Pairing} we plot the pairing energy for the equilibrium ground-state, the fission isomer, the top of the first 
and second barrier of $^{226}$Th. 
The pairing energies display a rapid decrease with temperature, and completely vanish beyond $T=0.75$ MeV. This is, of course,  
also the temperature at which the barrier heights start decreasing.

\begin{figure}
 \includegraphics[width=0.48\textwidth]{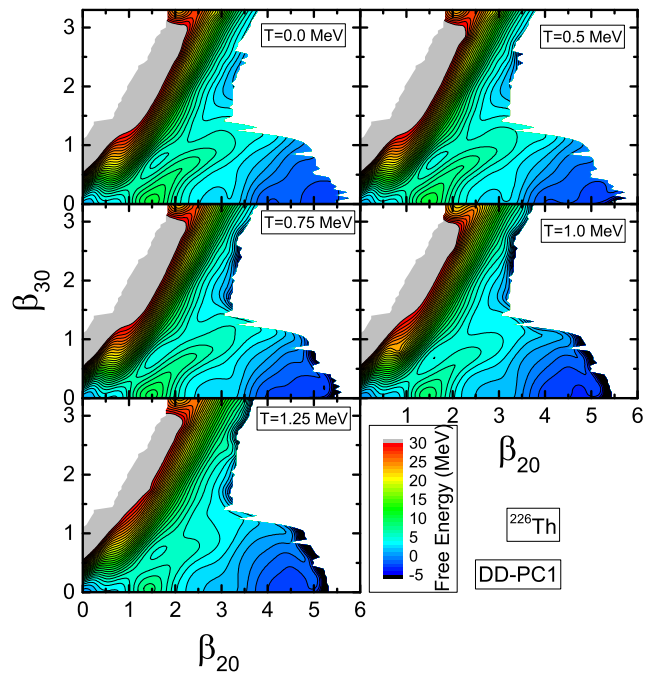}
\caption{(Color online)~\label{fig:Th226_FTPES}%
Free energy $F$ of $^{226}$Th in the $(\beta_{20},\beta_{30})$ plane for finite temperatures $T=0.0, 0.5, 0.75, 1.0, 1.25$ MeV.
In each panel energies are normalized with respect to the corresponding value at the equilibrium minimum, 
and contours join points on the surface with the same energy (in MeV). 
The energy surfaces are calculated with the relativistic density functionals DD-PC1~\cite{Niksic2008_PRC78-034318}, 
and the pairing interaction Eq.~(\ref{pairing}).
The contour interval is 1.0 MeV.
}
\end{figure}

The two-dimensional deformation free energy surfaces in the collective 
space $(\beta_{20},\beta_{30})$ at $T=0.0$, $0.5$, $0.75$, $1.0$, and $1.25$ MeV 
are shown in Fig.~\ref{fig:Th226_FTPES}. 
Only configurations with $\hat{Q}_{N} \geq 2$ are plotted, and 
the frontier of this domain determines the scission contour. 
The deformation surfaces at $T=0.0$ and $0.5$ are almost indistinguishable. 
The ridge separating the asymmetric and symmetric fission valleys gradually decreases with temperature for $T  \geq 0.75$ MeV.
The scission contour at various temperatures displays similar patterns, that is, it starts from an 
elongated symmetric point at $\beta_{20}  \sim 5.5$, 
and evolves to a minimal elongation $\beta_{20} \sim 3.0$ as asymmetry increases.
\begin{figure}
 \includegraphics[width=0.48\textwidth]{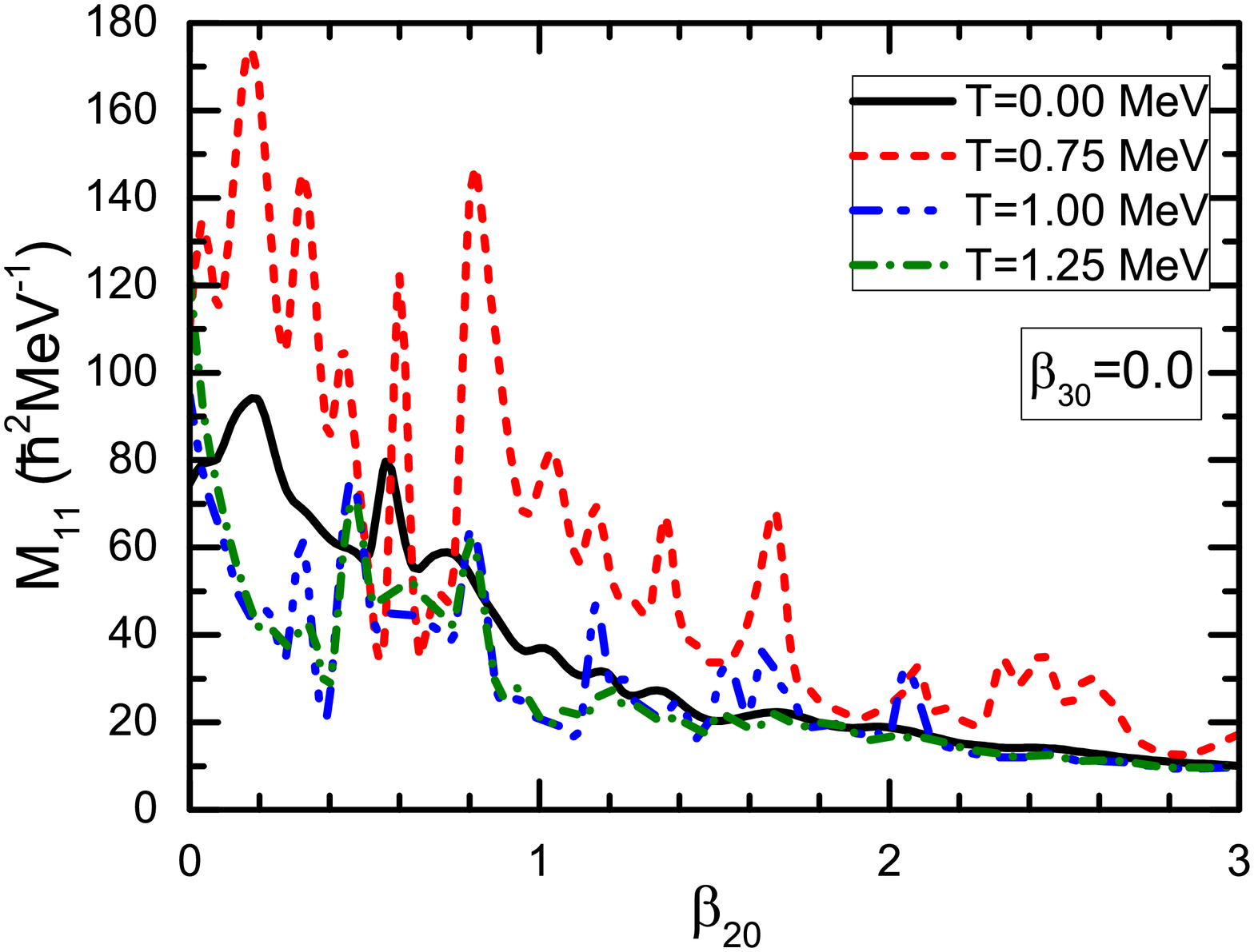}\\
  \includegraphics[width=0.48\textwidth]{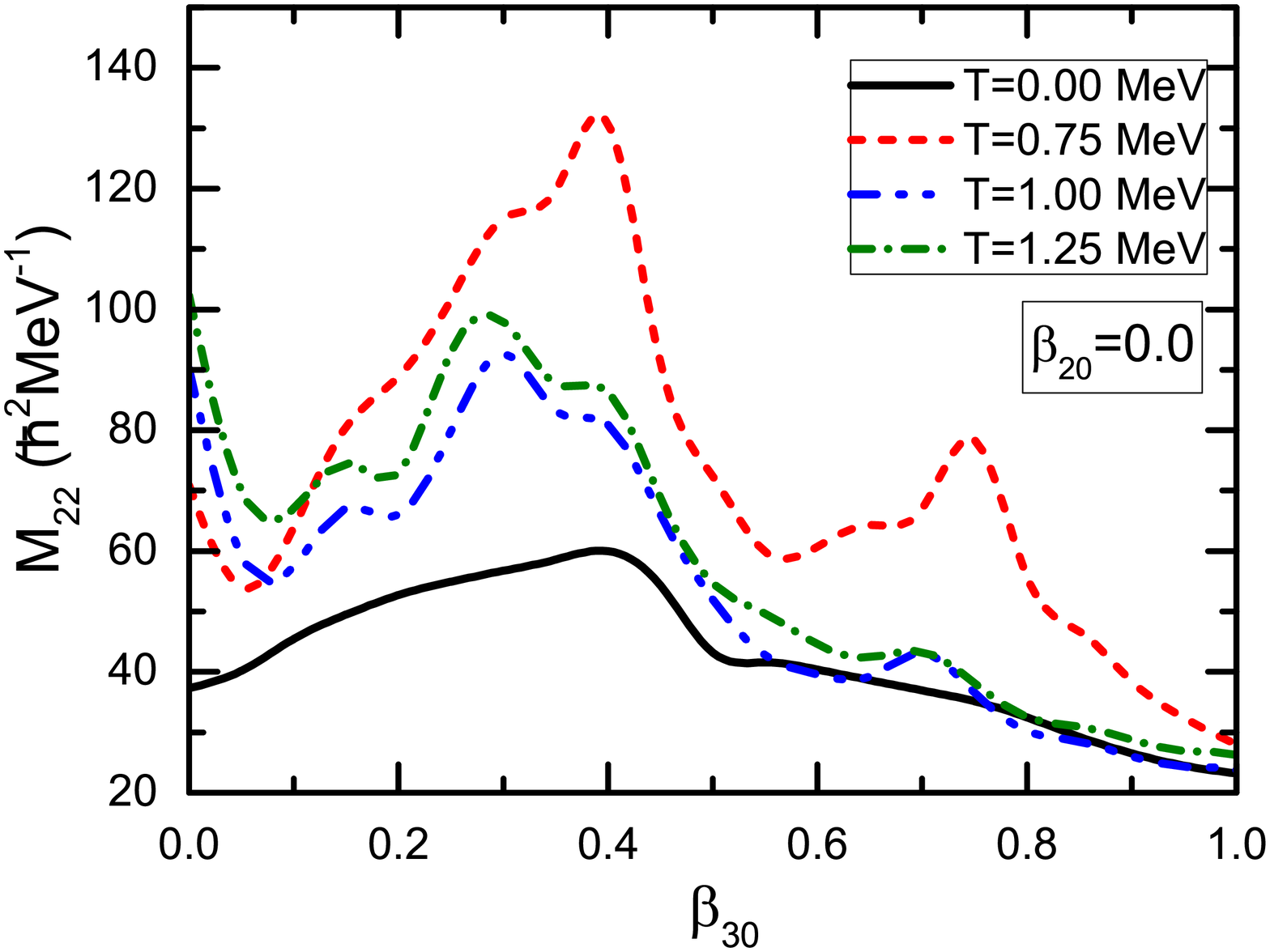}
\caption{(Color online)~\label{fig:Th226_TTMass}%
The $\mathcal{M}_{11}$ component of the inertia tensor of $^{226}$Th as function of the 
quadrupole deformation $\beta_{20}$ (upper panel), and 
the $\mathcal{M}_{22}$ component as function of the octupole deformation $\beta_{30}$ (lower panel) 
for finite temperatures $T=0.0, 0.75, 1.0$, and $1.25$ MeV. 
}
\end{figure}

For the two-dimensional space of collective deformation coordinates three independent components $\mathcal{M}_{11}$, 
$\mathcal{M}_{12}$, and $\mathcal{M}_{22}$ determine the inertia tensor. 
In the present case the indices $1$ and $2$ refer to the $\beta_{20}$ and $\beta_{30}$ degrees of freedom, respectively. 
In Fig.~\ref{fig:Th226_TTMass} the evolution of the $\mathcal{M}_{11}$ component of the collective mass with the quadrupole 
deformation parameter $\beta_{20}$, and the $\mathcal{M}_{22}$ component as function of the octupole deformation $\beta_{30}$, 
are shown for different temperatures. One first notices that $\mathcal{M}_{11}$ exhibits more oscillations that reflect the complex 
underlying structure of level crossings, while $\mathcal{M}_{22}$ displays a smooth behavior as a function of octupole deformation 
at $T=0$. In the interval $T=0 \sim 0.75$ MeV both components generally increase with temperature, 
due to the weakening of pairing correlations and reduction of pairing gaps for $T > 0$ MeV. Note that
in the first approximation the effective collective inertia $\mathcal{M} \propto \Delta^{-2}$, 
where $\Delta$ is the pairing gap~\cite{Moretto1974_PLB49-147}. After the 
pairing phase transition has occurred $\mathcal{M}_{11}$ and $\mathcal{M}_{22}$ decrease as a consequence 
of the weakening of shell effects, except for rather large values 
at the spherical shape. A similar behavior was also observed in studies based on non-relativistic
Skyrme \cite{Zhu2016_PRC94-024329} and Gogny functionals \cite{Martin2009_IJMPE18-861}.

\subsection{\label{subsec:yields}Evolution of charge and mass fragment distributions with temperature}

\begin{figure}
 \includegraphics[width=0.48\textwidth]{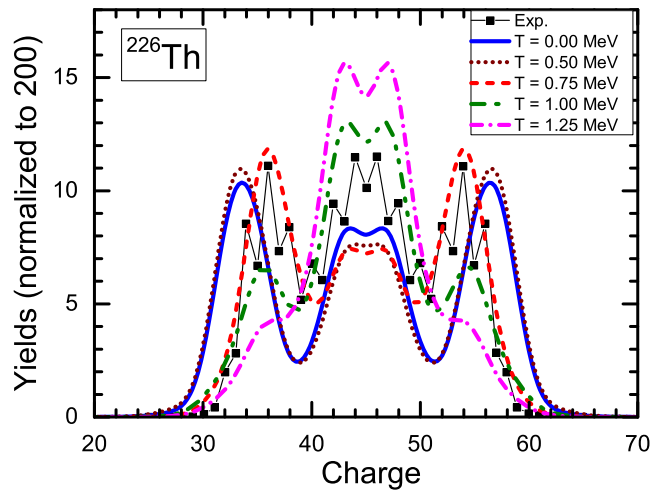}
\caption{(Color online)~\label{fig:Yields_Charge}%
Charge yields for induced fission of $^{226}$Th.
The collective potentials and perturbative cranking inertia tensors for the finite 
temperatures $T=0.0, 0.5, 0.75, 1.00, 1.25$ MeV 
are used in the calculations.  The corresponding internal excitation energies are 
$E_{\rm int.}^{*} = 0.0$, $2.58$, $8.71$, $16.56$, and $27.12$ MeV, respectively.
The average excitation energy of the initial state ($E_{\rm coll.}^{*}$) is chosen 1 MeV 
above the corresponding second fission barrier $B_{II}$.
The experimental charge yields for $^{226}$Th$(\gamma, f)$ are from  
Ref.~\cite{Schmidt2000_NPA665-221}.
}
\end{figure}

The dynamics of induced fission of $^{226}$Th at different temperatures is explored using the time-dependent 
generator coordinate method (TDGCM) in the Gaussian overlap approximation (GOA).
The potential entering the collective Hamiltonian Eq.~(\ref{eq:Hcoll}) is given by the Helmholtz free energy 
$F=E(T)-TS$, with $E(T)$ the RMF+BCS deformation energy in the $(\beta_{20},\beta_{30})$ plane, and the inertia tensor 
is calculated using Eq.~(\ref{eq:per-mass}). The average energy of the initial state $E_{\rm coll.}^{*}$  
is chosen 1 MeV above the corresponding second (higher) fission barrier $B_{II}$.

\begin{figure}
 \includegraphics[width=0.48\textwidth]{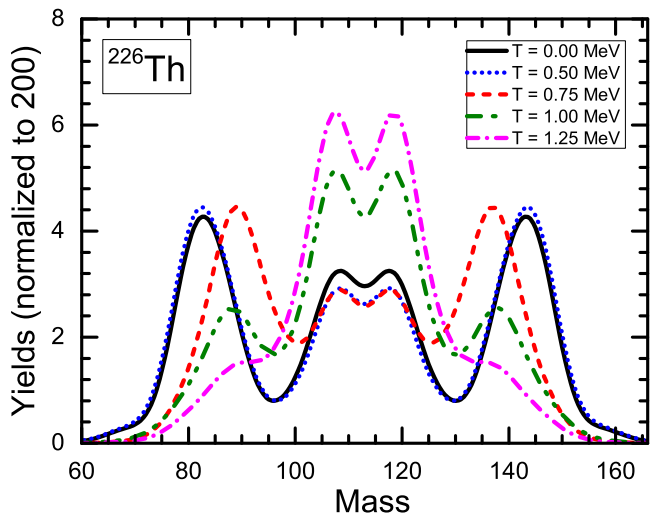}
\caption{(Color online)~\label{fig:Yields_Mass}%
Same as Fig.~\ref{fig:Yields_Charge}, but for pre-neutron emission mass yields.
}
\end{figure}

The pre-neutron emission charge and mass yields obtained with the TDGCM+GOA, and normalized to $\sum_{A} Y(A) = 200$, are shown in 
Figs.~\ref{fig:Yields_Charge} and~\ref{fig:Yields_Mass}, respectively.
The experimental fragment charge distribution of $^{226}$Th~\cite{Schmidt2000_NPA665-221} is also included in the plot of Fig.~\ref{fig:Yields_Charge}. For $T=0$ MeV the calculation reproduces the trend of the data except, of course, the odd-even staggering. 
In more detail, however, the predicted asymmetric peaks are located at $Z=34$ and $Z=56$, two mass units away from the experimental 
asymmetric peaks at $Z=36$ and $Z=54$. The empirical yield for 
symmetric fission is somewhat underestimated in the zero-temperature calculation. This picture does not change quantitatively for 
$T=0.5$ MeV, as this temperature corresponds to an internal excitation energy of only $E_{\rm int.}^{*}=2.58$ MeV and, therefore, 
the collective potential and inertia tensor are not modified significantly (cf. Sec.~\ref{subsec:barriers}).

At temperature $T=0.75$ MeV the asymmetric peaks of the charge yields are predicted at $Z=36$ and $Z=54$, 
in excellent agreement with the empirical values. However, the symmetric fission peak is still lower than the experimental one.
The corresponding internal excitation energy of the nucleus is $E_{\rm int.}^{*}=8.71$ MeV.
With a further increases of the temperature to $T=1.0$ MeV, corresponding to $E_{\rm int.}^{*}=16.56$ MeV, 
the yields of the asymmetric peaks at $Z=36$ and $Z=54$ decrease, whereas the symmetric peak increases above the 
experimental value. This can in part be attributed to the 
decreases of the ridge separating the asymmetric and symmetric fission valleys, as shown in Fig.~\ref{fig:Th226_FTPES}.
It is interesting to note that the experimental charge yield distribution lies between our theoretical results obtained 
for $E_{\rm int.}^{*}=8.71$ and $16.56$ MeV. Indeed, the experimental results were obtained in photoinduced fission with photon energies 
in the interval $8-14$ MeV, with a peak value of $E_{\gamma} = 11$ MeV~\cite{Schmidt2000_NPA665-221}.
Finally, the calculated charge distribution becomes almost completely symmetric at the highest temperature considered in the 
present study: $T=1.25$ MeV, corresponding to an internal excitation energy of $E_{\rm int.}^{*}=27.12$ MeV.

The calculated pre-neutron emission mass yields for different temperatures are shown in Fig.~\ref{fig:Yields_Mass}.
Analogous to the charge distributions shown in Fig.~\ref{fig:Yields_Charge}, a three-peak structure is obtained with the 
asymmetric peaks located at $A=83$ and $A=143$, for $T=0.0$ and $0.5$ MeV. 
At $T=0.75$ MeV the asymmetric peaks shift by 6 mass units to $A=89$ and $A=137$. With a 
further increases of temperature the yields of the asymmetric peaks decrease, while the symmetric-fission peak is enhanced.
The calculated distribution becomes symmetric at $T=1.25$ MeV.

\section{\label{sec:summary}Summary}

We have explored the dynamics of induced fission of $^{226}$Th in a theoretical framework based on the finite-temperature 
time-dependent generator coordinate method (TDGCM) in the Gaussian overlap approximation (GOA). The collective 
Hamiltonian is determined by the 
temperature-dependent free energy and perturbative cranking inertia tensor in the two dimensional 
space of quadrupole and octupole deformations $(\beta_{20},\beta_{30})$, calculated using the finite-temperature multidimensionally-constrained 
relativistic mean-field plus BCS model. The relativistic energy density functional DD-PC1 has been employed in the particle-hole 
channel, and pairing correlations treated in the BCS approximation using a separable pairing force of finite range. The TDGCM+GOA is used to 
propagate the initial collective state in time and describe fission dynamics.

The critical temperature for the pairing phase transition of $^{226}$Th is at $T \approx 0.75$ MeV.
At lower temperatures one notices only small changes in the potential (free) energy surface, while 
the inertia increases because of the weakening of pairing correlations. 
The fission barriers start to decrease at $T > 0.75$ MeV, as well as the ridge separating the symmetric and 
asymmetric fission valleys. The components of the inertia tensor decrease after the pairing phase transition. 

The pre-neutron emission charge and mass distributions are calculated using the FELIX code -- version 2.0, 
which is the most recent implementation of the TDGCM+GOA model.
Although the empirical triple-humped structure of the fission charge and mass distributions can qualitatively 
be described without taking into account temperature effects, the experimental positions of the asymmetric peaks 
and the symmetric-fission yield can only be accurately reproduced in the TDGCM+GOA by using the finite-temperature 
collective potential and inertia tensor. The model predicts a transition from asymmetric to symmetric fission of $^{226}$Th 
as the internal excitation energy increases. The charge and mass distributions are determined by the 
collective potential and inertia tensor, thus sensitive to the internal excitation energies of the compound nucleus, 
while the total flux as a function of time is more sensitive to the energy of the collective initial state. 

\bigskip
\acknowledgements
This work has been undertaken as part of the Inter-Governmental S\&T Cooperation Project between China and Croatia. 
It has also been supported in part by the QuantiXLie Centre of Excellence, a project co-financed by the Croatian Government and European Union through the European Regional Development Fund - the Competitiveness and Cohesion Operational Programme (KK.01.1.1.01).
Calculations have been performed in part at the HPC Cluster of KLTP/ITP-CAS and the Supercomputing Center, CNIC of CAS. S.G.Z. was supported by the National Key R\&D Program of China (2018YFA0404402), the NSF of China (11525524, 11621131001, 11647601, 11747601, and 11711540016), the CAS Key Research Program of Frontier Sciences (QYZDB-SSWSYS013), the CAS Key Research Program (XDPB09), and the IAEA CRP ``F41033''. 


\begin{thebibliography}{60}%
\makeatletter
\providecommand \@ifxundefined [1]{%
 \@ifx{#1\undefined}
}%
\providecommand \@ifnum [1]{%
 \ifnum #1\expandafter \@firstoftwo
 \else \expandafter \@secondoftwo
 \fi
}%
\providecommand \@ifx [1]{%
 \ifx #1\expandafter \@firstoftwo
 \else \expandafter \@secondoftwo
 \fi
}%
\providecommand \natexlab [1]{#1}%
\providecommand \enquote  [1]{``#1''}%
\providecommand \bibnamefont  [1]{#1}%
\providecommand \bibfnamefont [1]{#1}%
\providecommand \citenamefont [1]{#1}%
\providecommand \href@noop [0]{\@secondoftwo}%
\providecommand \href [0]{\begingroup \@sanitize@url \@href}%
\providecommand \@href[1]{\@@startlink{#1}\@@href}%
\providecommand \@@href[1]{\endgroup#1\@@endlink}%
\providecommand \@sanitize@url [0]{\catcode `\\12\catcode `\$12\catcode
  `\&12\catcode `\#12\catcode `\^12\catcode `\_12\catcode `\%12\relax}%
\providecommand \@@startlink[1]{}%
\providecommand \@@endlink[0]{}%
\providecommand \url  [0]{\begingroup\@sanitize@url \@url }%
\providecommand \@url [1]{\endgroup\@href {#1}{\urlprefix }}%
\providecommand \urlprefix  [0]{URL }%
\providecommand \Eprint [0]{\href }%
\providecommand \doibase [0]{http://dx.doi.org/}%
\providecommand \selectlanguage [0]{\@gobble}%
\providecommand \bibinfo  [0]{\@secondoftwo}%
\providecommand \bibfield  [0]{\@secondoftwo}%
\providecommand \translation [1]{[#1]}%
\providecommand \BibitemOpen [0]{}%
\providecommand \bibitemStop [0]{}%
\providecommand \bibitemNoStop [0]{.\EOS\space}%
\providecommand \EOS [0]{\spacefactor3000\relax}%
\providecommand \BibitemShut  [1]{\csname bibitem#1\endcsname}%
\let\auto@bib@innerbib\@empty
\bibitem [{\citenamefont {Schunck}\ and\ \citenamefont
  {Robledo}(2016)}]{Schunck2016_RPP79-116301}%
  \BibitemOpen
  \bibfield  {author} {\bibinfo {author} {\bibfnamefont {N.}~\bibnamefont
  {Schunck}}\ and\ \bibinfo {author} {\bibfnamefont {L.~M.}\ \bibnamefont
  {Robledo}},\ }\href {http://stacks.iop.org/0034-4885/79/i=11/a=116301}
  {\bibfield  {journal} {\bibinfo  {journal} {Rep. Prog. Phys.}\ }\textbf
  {\bibinfo {volume} {79}},\ \bibinfo {pages} {116301} (\bibinfo {year}
  {2016})}\BibitemShut {NoStop}%
\bibitem [{\citenamefont {Schmidt}\ \emph {et~al.}(2000)\citenamefont
  {Schmidt}, \citenamefont {Steinhauser}, \citenamefont {Bockstiegel},
  \citenamefont {Grewe}, \citenamefont {Heinz}, \citenamefont {Junghans},
  \citenamefont {Benlliure}, \citenamefont {Clerc}, \citenamefont {de~Jong},
  \citenamefont {Muller}, \citenamefont {Pfutzner},\ and\ \citenamefont
  {Voss}}]{Schmidt2000_NPA665-221}%
  \BibitemOpen
  \bibfield  {author} {\bibinfo {author} {\bibfnamefont {K.-H.}\ \bibnamefont
  {Schmidt}}, \bibinfo {author} {\bibfnamefont {S.}~\bibnamefont
  {Steinhauser}}, \bibinfo {author} {\bibfnamefont {C.}~\bibnamefont
  {Bockstiegel}}, \bibinfo {author} {\bibfnamefont {A.}~\bibnamefont {Grewe}},
  \bibinfo {author} {\bibfnamefont {A.}~\bibnamefont {Heinz}}, \bibinfo
  {author} {\bibfnamefont {A.}~\bibnamefont {Junghans}}, \bibinfo {author}
  {\bibfnamefont {J.}~\bibnamefont {Benlliure}}, \bibinfo {author}
  {\bibfnamefont {H.-G.}\ \bibnamefont {Clerc}}, \bibinfo {author}
  {\bibfnamefont {M.}~\bibnamefont {de~Jong}}, \bibinfo {author} {\bibfnamefont
  {J.}~\bibnamefont {Muller}}, \bibinfo {author} {\bibfnamefont
  {M.}~\bibnamefont {Pfutzner}}, \ and\ \bibinfo {author} {\bibfnamefont
  {B.}~\bibnamefont {Voss}},\ }\href
  {http://www.sciencedirect.com/science/article/pii/S037594749900384X}
  {\bibfield  {journal} {\bibinfo  {journal} {Nucl. Phys. A}\ }\textbf
  {\bibinfo {volume} {665}},\ \bibinfo {pages} {221} (\bibinfo {year}
  {2000})}\BibitemShut {NoStop}%
\bibitem [{\citenamefont {Andreyev}\ \emph {et~al.}(2013)\citenamefont
  {Andreyev}, \citenamefont {Huyse},\ and\ \citenamefont
  {Van~Duppen}}]{Andreyev2013_RMP85-1541}%
  \BibitemOpen
  \bibfield  {author} {\bibinfo {author} {\bibfnamefont {A.~N.}\ \bibnamefont
  {Andreyev}}, \bibinfo {author} {\bibfnamefont {M.}~\bibnamefont {Huyse}}, \
  and\ \bibinfo {author} {\bibfnamefont {P.}~\bibnamefont {Van~Duppen}},\
  }\href {http://link.aps.org/doi/10.1103/RevModPhys.85.1541} {\bibfield
  {journal} {\bibinfo  {journal} {Rev. Mod. Phys.}\ }\textbf {\bibinfo {volume}
  {85}},\ \bibinfo {pages} {1541} (\bibinfo {year} {2013})}\BibitemShut
  {NoStop}%
\bibitem [{\citenamefont {Isaev}\ \emph {et~al.}(2008)\citenamefont {Isaev},
  \citenamefont {Prieels}, \citenamefont {Keutgen}, \citenamefont {Van~Mol},
  \citenamefont {El~Masri},\ and\ \citenamefont
  {Demetriou}}]{Isaev2008_NPA809-1}%
  \BibitemOpen
  \bibfield  {author} {\bibinfo {author} {\bibfnamefont {S.}~\bibnamefont
  {Isaev}}, \bibinfo {author} {\bibfnamefont {R.}~\bibnamefont {Prieels}},
  \bibinfo {author} {\bibfnamefont {T.}~\bibnamefont {Keutgen}}, \bibinfo
  {author} {\bibfnamefont {J.}~\bibnamefont {Van~Mol}}, \bibinfo {author}
  {\bibfnamefont {Y.}~\bibnamefont {El~Masri}}, \ and\ \bibinfo {author}
  {\bibfnamefont {P.}~\bibnamefont {Demetriou}},\ }\href
  {http://www.sciencedirect.com/science/article/pii/S0375947408005708}
  {\bibfield  {journal} {\bibinfo  {journal} {Nucl. Phys. A}\ }\textbf
  {\bibinfo {volume} {809}},\ \bibinfo {pages} {1} (\bibinfo {year}
  {2008})}\BibitemShut {NoStop}%
\bibitem [{\citenamefont {Ryzhov}\ \emph {et~al.}(2011)\citenamefont {Ryzhov},
  \citenamefont {Yavshits}, \citenamefont {Tutin}, \citenamefont {Kovalev},
  \citenamefont {Saulski}, \citenamefont {Kudryashev}, \citenamefont {Onegin},
  \citenamefont {Vaishnene}, \citenamefont {Gavrikov}, \citenamefont
  {Grudzevich}, \citenamefont {Simutkin}, \citenamefont {Pomp}, \citenamefont
  {Blomgren}, \citenamefont {\&Ouml;sterlund}, \citenamefont {Andersson},
  \citenamefont {Bevilacqua}, \citenamefont {Meulders},\ and\ \citenamefont
  {Prieels}}]{Ryzhov2011_PRC83-054603}%
  \BibitemOpen
  \bibfield  {author} {\bibinfo {author} {\bibfnamefont {I.~V.}\ \bibnamefont
  {Ryzhov}}, \bibinfo {author} {\bibfnamefont {S.~G.}\ \bibnamefont
  {Yavshits}}, \bibinfo {author} {\bibfnamefont {G.~A.}\ \bibnamefont {Tutin}},
  \bibinfo {author} {\bibfnamefont {N.~V.}\ \bibnamefont {Kovalev}}, \bibinfo
  {author} {\bibfnamefont {A.~V.}\ \bibnamefont {Saulski}}, \bibinfo {author}
  {\bibfnamefont {N.~A.}\ \bibnamefont {Kudryashev}}, \bibinfo {author}
  {\bibfnamefont {M.~S.}\ \bibnamefont {Onegin}}, \bibinfo {author}
  {\bibfnamefont {L.~A.}\ \bibnamefont {Vaishnene}}, \bibinfo {author}
  {\bibfnamefont {Y.~A.}\ \bibnamefont {Gavrikov}}, \bibinfo {author}
  {\bibfnamefont {O.~T.}\ \bibnamefont {Grudzevich}}, \bibinfo {author}
  {\bibfnamefont {V.~D.}\ \bibnamefont {Simutkin}}, \bibinfo {author}
  {\bibfnamefont {S.}~\bibnamefont {Pomp}}, \bibinfo {author} {\bibfnamefont
  {J.}~\bibnamefont {Blomgren}}, \bibinfo {author} {\bibfnamefont
  {M.}~\bibnamefont {\&Ouml;sterlund}}, \bibinfo {author} {\bibfnamefont
  {P.}~\bibnamefont {Andersson}}, \bibinfo {author} {\bibfnamefont
  {R.}~\bibnamefont {Bevilacqua}}, \bibinfo {author} {\bibfnamefont {J.~P.}\
  \bibnamefont {Meulders}}, \ and\ \bibinfo {author} {\bibfnamefont
  {R.}~\bibnamefont {Prieels}},\ }\href
  {http://link.aps.org/doi/10.1103/PhysRevC.83.054603} {\bibfield  {journal}
  {\bibinfo  {journal} {Phys. Rev. C}\ }\textbf {\bibinfo {volume} {83}},\
  \bibinfo {pages} {054603} (\bibinfo {year} {2011})}\BibitemShut {NoStop}%
\bibitem [{\citenamefont {Sen}\ \emph {et~al.}(2017)\citenamefont {Sen},
  \citenamefont {Ghosh}, \citenamefont {Bhattacharya}, \citenamefont
  {Banerjee}, \citenamefont {Bhattacharya}, \citenamefont {Kundu},
  \citenamefont {Mukherjee}, \citenamefont {Asgar}, \citenamefont {Dey},
  \citenamefont {Dhal}, \citenamefont {Shaikh}, \citenamefont {Meena},
  \citenamefont {Manna}, \citenamefont {Pandey}, \citenamefont {Rana},
  \citenamefont {Roy}, \citenamefont {Roy}, \citenamefont {Srivastava},\ and\
  \citenamefont {Bhattacharya}}]{Sen2017_PRC96-064609}%
  \BibitemOpen
  \bibfield  {author} {\bibinfo {author} {\bibfnamefont {A.}~\bibnamefont
  {Sen}}, \bibinfo {author} {\bibfnamefont {T.~K.}\ \bibnamefont {Ghosh}},
  \bibinfo {author} {\bibfnamefont {S.}~\bibnamefont {Bhattacharya}}, \bibinfo
  {author} {\bibfnamefont {K.}~\bibnamefont {Banerjee}}, \bibinfo {author}
  {\bibfnamefont {C.}~\bibnamefont {Bhattacharya}}, \bibinfo {author}
  {\bibfnamefont {S.}~\bibnamefont {Kundu}}, \bibinfo {author} {\bibfnamefont
  {G.}~\bibnamefont {Mukherjee}}, \bibinfo {author} {\bibfnamefont
  {A.}~\bibnamefont {Asgar}}, \bibinfo {author} {\bibfnamefont
  {A.}~\bibnamefont {Dey}}, \bibinfo {author} {\bibfnamefont {A.}~\bibnamefont
  {Dhal}}, \bibinfo {author} {\bibfnamefont {M.~M.}\ \bibnamefont {Shaikh}},
  \bibinfo {author} {\bibfnamefont {J.~K.}\ \bibnamefont {Meena}}, \bibinfo
  {author} {\bibfnamefont {S.}~\bibnamefont {Manna}}, \bibinfo {author}
  {\bibfnamefont {R.}~\bibnamefont {Pandey}}, \bibinfo {author} {\bibfnamefont
  {T.~K.}\ \bibnamefont {Rana}}, \bibinfo {author} {\bibfnamefont
  {P.}~\bibnamefont {Roy}}, \bibinfo {author} {\bibfnamefont {T.}~\bibnamefont
  {Roy}}, \bibinfo {author} {\bibfnamefont {V.}~\bibnamefont {Srivastava}}, \
  and\ \bibinfo {author} {\bibfnamefont {P.}~\bibnamefont {Bhattacharya}},\
  }\href {https://link.aps.org/doi/10.1103/PhysRevC.96.064609} {\bibfield
  {journal} {\bibinfo  {journal} {Phys. Rev. C}\ }\textbf {\bibinfo {volume}
  {96}},\ \bibinfo {pages} {064609} (\bibinfo {year} {2017})}\BibitemShut
  {NoStop}%
\bibitem [{\citenamefont {Leguillon}\ \emph {et~al.}(2016)\citenamefont
  {Leguillon}, \citenamefont {Nishio}, \citenamefont {Hirose}, \citenamefont
  {Makii}, \citenamefont {Nishinaka}, \citenamefont {Orlandi}, \citenamefont
  {Tsukada}, \citenamefont {Smallcombe}, \citenamefont {Chiba}, \citenamefont
  {Aritomo}, \citenamefont {Ohtsuki}, \citenamefont {Tatsuzawa}, \citenamefont
  {Takaki}, \citenamefont {Tamura}, \citenamefont {Goto}, \citenamefont
  {Tsekhanovich}, \citenamefont {Petrache},\ and\ \citenamefont
  {Andreyev}}]{Leguillon2016_PLB761-125}%
  \BibitemOpen
  \bibfield  {author} {\bibinfo {author} {\bibfnamefont {R.}~\bibnamefont
  {Leguillon}}, \bibinfo {author} {\bibfnamefont {K.}~\bibnamefont {Nishio}},
  \bibinfo {author} {\bibfnamefont {K.}~\bibnamefont {Hirose}}, \bibinfo
  {author} {\bibfnamefont {H.}~\bibnamefont {Makii}}, \bibinfo {author}
  {\bibfnamefont {I.}~\bibnamefont {Nishinaka}}, \bibinfo {author}
  {\bibfnamefont {R.}~\bibnamefont {Orlandi}}, \bibinfo {author} {\bibfnamefont
  {K.}~\bibnamefont {Tsukada}}, \bibinfo {author} {\bibfnamefont
  {J.}~\bibnamefont {Smallcombe}}, \bibinfo {author} {\bibfnamefont
  {S.}~\bibnamefont {Chiba}}, \bibinfo {author} {\bibfnamefont
  {Y.}~\bibnamefont {Aritomo}}, \bibinfo {author} {\bibfnamefont
  {T.}~\bibnamefont {Ohtsuki}}, \bibinfo {author} {\bibfnamefont
  {R.}~\bibnamefont {Tatsuzawa}}, \bibinfo {author} {\bibfnamefont
  {N.}~\bibnamefont {Takaki}}, \bibinfo {author} {\bibfnamefont
  {N.}~\bibnamefont {Tamura}}, \bibinfo {author} {\bibfnamefont
  {S.}~\bibnamefont {Goto}}, \bibinfo {author} {\bibfnamefont {I.}~\bibnamefont
  {Tsekhanovich}}, \bibinfo {author} {\bibfnamefont {C.}~\bibnamefont
  {Petrache}}, \ and\ \bibinfo {author} {\bibfnamefont {A.}~\bibnamefont
  {Andreyev}},\ }\href
  {http://www.sciencedirect.com/science/article/pii/S0370269316304300}
  {\bibfield  {journal} {\bibinfo  {journal} {Phys. Lett. B}\ }\textbf
  {\bibinfo {volume} {761}},\ \bibinfo {pages} {125} (\bibinfo {year}
  {2016})}\BibitemShut {NoStop}%
\bibitem [{\citenamefont {Demetriou}\ and\ \citenamefont
  {Goriely}(2001)}]{Demetriou2001_NPA695-95}%
  \BibitemOpen
  \bibfield  {author} {\bibinfo {author} {\bibfnamefont {P.}~\bibnamefont
  {Demetriou}}\ and\ \bibinfo {author} {\bibfnamefont {S.}~\bibnamefont
  {Goriely}},\ }\href
  {http://www.sciencedirect.com/science/article/pii/S0375947401010958}
  {\bibfield  {journal} {\bibinfo  {journal} {Nucl. Phys. A}\ }\textbf
  {\bibinfo {volume} {695}},\ \bibinfo {pages} {95} (\bibinfo {year}
  {2001})}\BibitemShut {NoStop}%
\bibitem [{\citenamefont {Hirose}\ \emph {et~al.}(2017)\citenamefont {Hirose},
  \citenamefont {Nishio}, \citenamefont {Tanaka}, \citenamefont {Leguillon},
  \citenamefont {Makii}, \citenamefont {Nishinaka}, \citenamefont {Orlandi},
  \citenamefont {Tsukada}, \citenamefont {Smallcombe}, \citenamefont
  {Vermeulen}, \citenamefont {Chiba}, \citenamefont {Aritomo}, \citenamefont
  {Ohtsuki}, \citenamefont {Nakano}, \citenamefont {Araki}, \citenamefont
  {Watanabe}, \citenamefont {Tatsuzawa}, \citenamefont {Takaki}, \citenamefont
  {Tamura}, \citenamefont {Goto}, \citenamefont {Tsekhanovich},\ and\
  \citenamefont {Andreyev}}]{Hirose2017_PRL119-222501}%
  \BibitemOpen
  \bibfield  {author} {\bibinfo {author} {\bibfnamefont {K.}~\bibnamefont
  {Hirose}}, \bibinfo {author} {\bibfnamefont {K.}~\bibnamefont {Nishio}},
  \bibinfo {author} {\bibfnamefont {S.}~\bibnamefont {Tanaka}}, \bibinfo
  {author} {\bibfnamefont {R.}~\bibnamefont {Leguillon}}, \bibinfo {author}
  {\bibfnamefont {H.}~\bibnamefont {Makii}}, \bibinfo {author} {\bibfnamefont
  {I.}~\bibnamefont {Nishinaka}}, \bibinfo {author} {\bibfnamefont
  {R.}~\bibnamefont {Orlandi}}, \bibinfo {author} {\bibfnamefont
  {K.}~\bibnamefont {Tsukada}}, \bibinfo {author} {\bibfnamefont
  {J.}~\bibnamefont {Smallcombe}}, \bibinfo {author} {\bibfnamefont {M.~J.}\
  \bibnamefont {Vermeulen}}, \bibinfo {author} {\bibfnamefont {S.}~\bibnamefont
  {Chiba}}, \bibinfo {author} {\bibfnamefont {Y.}~\bibnamefont {Aritomo}},
  \bibinfo {author} {\bibfnamefont {T.}~\bibnamefont {Ohtsuki}}, \bibinfo
  {author} {\bibfnamefont {K.}~\bibnamefont {Nakano}}, \bibinfo {author}
  {\bibfnamefont {S.}~\bibnamefont {Araki}}, \bibinfo {author} {\bibfnamefont
  {Y.}~\bibnamefont {Watanabe}}, \bibinfo {author} {\bibfnamefont
  {R.}~\bibnamefont {Tatsuzawa}}, \bibinfo {author} {\bibfnamefont
  {N.}~\bibnamefont {Takaki}}, \bibinfo {author} {\bibfnamefont
  {N.}~\bibnamefont {Tamura}}, \bibinfo {author} {\bibfnamefont
  {S.}~\bibnamefont {Goto}}, \bibinfo {author} {\bibfnamefont {I.}~\bibnamefont
  {Tsekhanovich}}, \ and\ \bibinfo {author} {\bibfnamefont {A.~N.}\
  \bibnamefont {Andreyev}},\ }\href
  {https://link.aps.org/doi/10.1103/PhysRevLett.119.222501} {\bibfield
  {journal} {\bibinfo  {journal} {Phys. Rev. Lett.}\ }\textbf {\bibinfo
  {volume} {119}},\ \bibinfo {pages} {222501} (\bibinfo {year}
  {2017})}\BibitemShut {NoStop}%
\bibitem [{\citenamefont {Simutkin}\ \emph {et~al.}(2014)\citenamefont
  {Simutkin}, \citenamefont {Pomp}, \citenamefont {Blomgren}, \citenamefont
  {\"Osterlund}, \citenamefont {Bevilacqua}, \citenamefont {Andersson},
  \citenamefont {Ryzhov}, \citenamefont {Tutin}, \citenamefont {Yavshits},
  \citenamefont {Vaishnene}, \citenamefont {Onegin}, \citenamefont {Meulders},\
  and\ \citenamefont {Prieels}}]{Simutkin2014_NDS119-331}%
  \BibitemOpen
  \bibfield  {author} {\bibinfo {author} {\bibfnamefont {V.}~\bibnamefont
  {Simutkin}}, \bibinfo {author} {\bibfnamefont {S.}~\bibnamefont {Pomp}},
  \bibinfo {author} {\bibfnamefont {J.}~\bibnamefont {Blomgren}}, \bibinfo
  {author} {\bibfnamefont {M.}~\bibnamefont {\"Osterlund}}, \bibinfo {author}
  {\bibfnamefont {R.}~\bibnamefont {Bevilacqua}}, \bibinfo {author}
  {\bibfnamefont {P.}~\bibnamefont {Andersson}}, \bibinfo {author}
  {\bibfnamefont {I.}~\bibnamefont {Ryzhov}}, \bibinfo {author} {\bibfnamefont
  {G.}~\bibnamefont {Tutin}}, \bibinfo {author} {\bibfnamefont
  {S.}~\bibnamefont {Yavshits}}, \bibinfo {author} {\bibfnamefont
  {L.}~\bibnamefont {Vaishnene}}, \bibinfo {author} {\bibfnamefont
  {M.}~\bibnamefont {Onegin}}, \bibinfo {author} {\bibfnamefont
  {J.}~\bibnamefont {Meulders}}, \ and\ \bibinfo {author} {\bibfnamefont
  {R.}~\bibnamefont {Prieels}},\ }\href
  {http://www.sciencedirect.com/science/article/pii/S0090375214006255}
  {\bibfield  {journal} {\bibinfo  {journal} {Nucl. Data Sheets}\ }\textbf
  {\bibinfo {volume} {119}},\ \bibinfo {pages} {331} (\bibinfo {year}
  {2014})}\BibitemShut {NoStop}%
\bibitem [{\citenamefont {Ramos}\ \emph {et~al.}(2018)\citenamefont {Ramos},
  \citenamefont {Caamano}, \citenamefont {Farget}, \citenamefont
  {Rodrlguez-Tajes}, \citenamefont {Audouin}, \citenamefont {Benlliure},
  \citenamefont {Casarejos}, \citenamefont {Clement}, \citenamefont {Cortina},
  \citenamefont {Delaune}, \citenamefont {Derkx}, \citenamefont {Dijon},
  \citenamefont {Dore}, \citenamefont {Fernandez-Domlnguez}, \citenamefont
  {de~France}, \citenamefont {Heinz}, \citenamefont {Jacquot}, \citenamefont
  {Navin}, \citenamefont {Paradela}, \citenamefont {Rejmund}, \citenamefont
  {Roger}, \citenamefont {Salsac},\ and\ \citenamefont
  {Schmitt}}]{Ramos2018_PRC97-054612}%
  \BibitemOpen
  \bibfield  {author} {\bibinfo {author} {\bibfnamefont {D.}~\bibnamefont
  {Ramos}}, \bibinfo {author} {\bibfnamefont {M.}~\bibnamefont {Caamano}},
  \bibinfo {author} {\bibfnamefont {F.}~\bibnamefont {Farget}}, \bibinfo
  {author} {\bibfnamefont {C.}~\bibnamefont {Rodrlguez-Tajes}}, \bibinfo
  {author} {\bibfnamefont {L.}~\bibnamefont {Audouin}}, \bibinfo {author}
  {\bibfnamefont {J.}~\bibnamefont {Benlliure}}, \bibinfo {author}
  {\bibfnamefont {E.}~\bibnamefont {Casarejos}}, \bibinfo {author}
  {\bibfnamefont {E.}~\bibnamefont {Clement}}, \bibinfo {author} {\bibfnamefont
  {D.}~\bibnamefont {Cortina}}, \bibinfo {author} {\bibfnamefont
  {O.}~\bibnamefont {Delaune}}, \bibinfo {author} {\bibfnamefont
  {X.}~\bibnamefont {Derkx}}, \bibinfo {author} {\bibfnamefont
  {A.}~\bibnamefont {Dijon}}, \bibinfo {author} {\bibfnamefont
  {D.}~\bibnamefont {Dore}}, \bibinfo {author} {\bibfnamefont {B.}~\bibnamefont
  {Fernandez-Domlnguez}}, \bibinfo {author} {\bibfnamefont {G.}~\bibnamefont
  {de~France}}, \bibinfo {author} {\bibfnamefont {A.}~\bibnamefont {Heinz}},
  \bibinfo {author} {\bibfnamefont {B.}~\bibnamefont {Jacquot}}, \bibinfo
  {author} {\bibfnamefont {A.}~\bibnamefont {Navin}}, \bibinfo {author}
  {\bibfnamefont {C.}~\bibnamefont {Paradela}}, \bibinfo {author}
  {\bibfnamefont {M.}~\bibnamefont {Rejmund}}, \bibinfo {author} {\bibfnamefont
  {T.}~\bibnamefont {Roger}}, \bibinfo {author} {\bibfnamefont {M.-D.}\
  \bibnamefont {Salsac}}, \ and\ \bibinfo {author} {\bibfnamefont
  {C.}~\bibnamefont {Schmitt}},\ }\href
  {https://link.aps.org/doi/10.1103/PhysRevC.97.054612} {\bibfield  {journal}
  {\bibinfo  {journal} {Phys. Rev. C}\ }\textbf {\bibinfo {volume} {97}},\
  \bibinfo {pages} {054612} (\bibinfo {year} {2018})}\BibitemShut {NoStop}%
\bibitem [{\citenamefont {Naik}\ \emph {et~al.}(2018)\citenamefont {Naik},
  \citenamefont {Kim},\ and\ \citenamefont {Kim}}]{Naik2018_PRC97-014614}%
  \BibitemOpen
  \bibfield  {author} {\bibinfo {author} {\bibfnamefont {H.}~\bibnamefont
  {Naik}}, \bibinfo {author} {\bibfnamefont {G.~N.}\ \bibnamefont {Kim}}, \
  and\ \bibinfo {author} {\bibfnamefont {K.}~\bibnamefont {Kim}},\ }\href
  {https://link.aps.org/doi/10.1103/PhysRevC.97.014614} {\bibfield  {journal}
  {\bibinfo  {journal} {Phys. Rev. C}\ }\textbf {\bibinfo {volume} {97}},\
  \bibinfo {pages} {014614} (\bibinfo {year} {2018})}\BibitemShut {NoStop}%
\bibitem [{\citenamefont {Berger}\ \emph {et~al.}(1991)\citenamefont {Berger},
  \citenamefont {Girod},\ and\ \citenamefont {Gogny}}]{Berger1991_CPC63-365}%
  \BibitemOpen
  \bibfield  {author} {\bibinfo {author} {\bibfnamefont {J.}~\bibnamefont
  {Berger}}, \bibinfo {author} {\bibfnamefont {M.}~\bibnamefont {Girod}}, \
  and\ \bibinfo {author} {\bibfnamefont {D.}~\bibnamefont {Gogny}},\ }\href
  {http://www.sciencedirect.com/science/article/pii/001046559190263K}
  {\bibfield  {journal} {\bibinfo  {journal} {Comput. Phys. Commun.}\ }\textbf
  {\bibinfo {volume} {63}},\ \bibinfo {pages} {365} (\bibinfo {year}
  {1991})}\BibitemShut {NoStop}%
\bibitem [{\citenamefont {Goutte}\ \emph {et~al.}(2005)\citenamefont {Goutte},
  \citenamefont {Berger}, \citenamefont {Casoli},\ and\ \citenamefont
  {Gogny}}]{Goutte2005_PRC71-024316}%
  \BibitemOpen
  \bibfield  {author} {\bibinfo {author} {\bibfnamefont {H.}~\bibnamefont
  {Goutte}}, \bibinfo {author} {\bibfnamefont {J.~F.}\ \bibnamefont {Berger}},
  \bibinfo {author} {\bibfnamefont {P.}~\bibnamefont {Casoli}}, \ and\ \bibinfo
  {author} {\bibfnamefont {D.}~\bibnamefont {Gogny}},\ }\href
  {http://link.aps.org/doi/10.1103/PhysRevC.71.024316} {\bibfield  {journal}
  {\bibinfo  {journal} {Phys. Rev. C}\ }\textbf {\bibinfo {volume} {71}},\
  \bibinfo {pages} {024316} (\bibinfo {year} {2005})}\BibitemShut {NoStop}%
\bibitem [{\citenamefont {Younes}\ and\ \citenamefont
  {Gogny}(2012)}]{Younes2012_LLNL-TR-586678}%
  \BibitemOpen
  \bibfield  {author} {\bibinfo {author} {\bibfnamefont {W.}~\bibnamefont
  {Younes}}\ and\ \bibinfo {author} {\bibfnamefont {D.}~\bibnamefont {Gogny}},\
  }\href@noop {} {\emph {\bibinfo {title} {Fragment Yields Calculated in a
  Time-Dependent Microscopic Theory of Fission}}},\ \bibinfo {type} {Tech.
  Rep.}\ (\bibinfo  {institution} {Lawrence Livermore National Laboratory},\
  \bibinfo {year} {2012})\BibitemShut {NoStop}%
\bibitem [{\citenamefont {Regnier}\ \emph {et~al.}(2018)\citenamefont
  {Regnier}, \citenamefont {Dubray}, \citenamefont {Verriere},\ and\
  \citenamefont {Schunck}}]{Regnier2018_CPC225-180}%
  \BibitemOpen
  \bibfield  {author} {\bibinfo {author} {\bibfnamefont {D.}~\bibnamefont
  {Regnier}}, \bibinfo {author} {\bibfnamefont {N.}~\bibnamefont {Dubray}},
  \bibinfo {author} {\bibfnamefont {M.}~\bibnamefont {Verriere}}, \ and\
  \bibinfo {author} {\bibfnamefont {N.}~\bibnamefont {Schunck}},\ }\href
  {https://www.sciencedirect.com/science/article/pii/S0010465517304125}
  {\bibfield  {journal} {\bibinfo  {journal} {Comput. Phys. Commun.}\ }\textbf
  {\bibinfo {volume} {225}},\ \bibinfo {pages} {180} (\bibinfo {year}
  {2018})}\BibitemShut {NoStop}%
\bibitem [{\citenamefont {Regnier}\ \emph {et~al.}(2017)\citenamefont
  {Regnier}, \citenamefont {Dubray}, \citenamefont {Schunck},\ and\
  \citenamefont {Verriere}}]{Regnier2017_EPJWC146-04043}%
  \BibitemOpen
  \bibfield  {author} {\bibinfo {author} {\bibfnamefont {D.}~\bibnamefont
  {Regnier}}, \bibinfo {author} {\bibfnamefont {N.}~\bibnamefont {Dubray}},
  \bibinfo {author} {\bibfnamefont {N.}~\bibnamefont {Schunck}}, \ and\
  \bibinfo {author} {\bibfnamefont {M.}~\bibnamefont {Verriere}},\ }\href
  {https://doi.org/10.1051/epjconf/201714604043} {\bibfield  {journal}
  {\bibinfo  {journal} {EPJ Web Conf.}\ }\textbf {\bibinfo {volume} {146}},\
  \bibinfo {pages} {04043} (\bibinfo {year} {2017})}\BibitemShut {NoStop}%
\bibitem [{\citenamefont {Regnier}\ \emph
  {et~al.}(2016{\natexlab{a}})\citenamefont {Regnier}, \citenamefont {Dubray},
  \citenamefont {Schunck},\ and\ \citenamefont
  {Verriere}}]{Regnier2016_PRC93-054611}%
  \BibitemOpen
  \bibfield  {author} {\bibinfo {author} {\bibfnamefont {D.}~\bibnamefont
  {Regnier}}, \bibinfo {author} {\bibfnamefont {N.}~\bibnamefont {Dubray}},
  \bibinfo {author} {\bibfnamefont {N.}~\bibnamefont {Schunck}}, \ and\
  \bibinfo {author} {\bibfnamefont {M.}~\bibnamefont {Verriere}},\ }\href
  {http://link.aps.org/doi/10.1103/PhysRevC.93.054611} {\bibfield  {journal}
  {\bibinfo  {journal} {Phys. Rev. C}\ }\textbf {\bibinfo {volume} {93}},\
  \bibinfo {pages} {054611} (\bibinfo {year} {2016}{\natexlab{a}})}\BibitemShut
  {NoStop}%
\bibitem [{\citenamefont {Regnier}\ \emph
  {et~al.}(2016{\natexlab{b}})\citenamefont {Regnier}, \citenamefont
  {Verriere}, \citenamefont {Dubray},\ and\ \citenamefont
  {Schunck}}]{Regnier2016_CPC200-350}%
  \BibitemOpen
  \bibfield  {author} {\bibinfo {author} {\bibfnamefont {D.}~\bibnamefont
  {Regnier}}, \bibinfo {author} {\bibfnamefont {M.}~\bibnamefont {Verriere}},
  \bibinfo {author} {\bibfnamefont {N.}~\bibnamefont {Dubray}}, \ and\ \bibinfo
  {author} {\bibfnamefont {N.}~\bibnamefont {Schunck}},\ }\href
  {http://www.sciencedirect.com/science/article/pii/S001046551500435X}
  {\bibfield  {journal} {\bibinfo  {journal} {Comput. Phys. Commun.}\ }\textbf
  {\bibinfo {volume} {200}},\ \bibinfo {pages} {350} (\bibinfo {year}
  {2016}{\natexlab{b}})}\BibitemShut {NoStop}%
\bibitem [{\citenamefont {Zdeb}\ \emph {et~al.}(2017)\citenamefont {Zdeb},
  \citenamefont {Dobrowolski},\ and\ \citenamefont
  {Warda}}]{Zdeb2017_PRC95-054608}%
  \BibitemOpen
  \bibfield  {author} {\bibinfo {author} {\bibfnamefont {A.}~\bibnamefont
  {Zdeb}}, \bibinfo {author} {\bibfnamefont {A.}~\bibnamefont {Dobrowolski}}, \
  and\ \bibinfo {author} {\bibfnamefont {M.}~\bibnamefont {Warda}},\ }\href
  {https://link.aps.org/doi/10.1103/PhysRevC.95.054608} {\bibfield  {journal}
  {\bibinfo  {journal} {Phys. Rev. C}\ }\textbf {\bibinfo {volume} {95}},\
  \bibinfo {pages} {054608} (\bibinfo {year} {2017})}\BibitemShut {NoStop}%
\bibitem [{\citenamefont {Vretenar}\ \emph {et~al.}(2005)\citenamefont
  {Vretenar}, \citenamefont {Afanasjev}, \citenamefont {Lalazissis},\ and\
  \citenamefont {Ring}}]{Vretenar2005_PR409-101}%
  \BibitemOpen
  \bibfield  {author} {\bibinfo {author} {\bibfnamefont {D.}~\bibnamefont
  {Vretenar}}, \bibinfo {author} {\bibfnamefont {A.}~\bibnamefont {Afanasjev}},
  \bibinfo {author} {\bibfnamefont {G.}~\bibnamefont {Lalazissis}}, \ and\
  \bibinfo {author} {\bibfnamefont {P.}~\bibnamefont {Ring}},\ }\href {\doibase
  DOI: 10.1016/j.physrep.2004.10.001} {\bibfield  {journal} {\bibinfo
  {journal} {Phys. Rep.}\ }\textbf {\bibinfo {volume} {409}},\ \bibinfo {pages}
  {101 } (\bibinfo {year} {2005})}\BibitemShut {NoStop}%
\bibitem [{\citenamefont {Meng}\ \emph {et~al.}(2006)\citenamefont {Meng},
  \citenamefont {Toki}, \citenamefont {Zhou}, \citenamefont {Zhang},
  \citenamefont {Long},\ and\ \citenamefont {Geng}}]{Meng2006_PPNP57-470}%
  \BibitemOpen
  \bibfield  {author} {\bibinfo {author} {\bibfnamefont {J.}~\bibnamefont
  {Meng}}, \bibinfo {author} {\bibfnamefont {H.}~\bibnamefont {Toki}}, \bibinfo
  {author} {\bibfnamefont {S.}~\bibnamefont {Zhou}}, \bibinfo {author}
  {\bibfnamefont {S.}~\bibnamefont {Zhang}}, \bibinfo {author} {\bibfnamefont
  {W.}~\bibnamefont {Long}}, \ and\ \bibinfo {author} {\bibfnamefont
  {L.}~\bibnamefont {Geng}},\ }\href {\doibase DOI: 10.1016/j.ppnp.2005.06.001}
  {\bibfield  {journal} {\bibinfo  {journal} {Prog. Part. Nucl. Phys.}\
  }\textbf {\bibinfo {volume} {57}},\ \bibinfo {pages} {470 } (\bibinfo {year}
  {2006})}\BibitemShut {NoStop}%
\bibitem [{\citenamefont {Meng}(2016)}]{Meng2016_WorldSci}%
  \BibitemOpen
  \bibinfo {editor} {\bibfnamefont {J.}~\bibnamefont {Meng}},\ ed.,\ \href@noop
  {} {\emph {\bibinfo {title} {Relativistic Density Functional for Nuclear
  Structure}}},\ \bibinfo {series} {International Review of Nuclear Physics},
  Vol.~\bibinfo {volume} {10}\ (\bibinfo  {publisher} {World Scientific},\
  \bibinfo {year} {2016})\BibitemShut {NoStop}%
\bibitem [{\citenamefont {Zhou}(2016)}]{Zhou2016_PS91-063008}%
  \BibitemOpen
  \bibfield  {author} {\bibinfo {author} {\bibfnamefont {S.-G.}\ \bibnamefont
  {Zhou}},\ }\href {\doibase 10.1088/0031-8949/91/6/063008} {\bibfield
  {journal} {\bibinfo  {journal} {Phys. Scr.}\ }\textbf {\bibinfo {volume}
  {91}},\ \bibinfo {pages} {063008} (\bibinfo {year} {2016})}\BibitemShut
  {NoStop}%
\bibitem [{\citenamefont {Burvenich}\ \emph {et~al.}(2004)\citenamefont
  {Burvenich}, \citenamefont {Bender}, \citenamefont {Maruhn},\ and\
  \citenamefont {Reinhard}}]{Burvenich2004_PRC69-014307}%
  \BibitemOpen
  \bibfield  {author} {\bibinfo {author} {\bibfnamefont {T.}~\bibnamefont
  {Burvenich}}, \bibinfo {author} {\bibfnamefont {M.}~\bibnamefont {Bender}},
  \bibinfo {author} {\bibfnamefont {J.~A.}\ \bibnamefont {Maruhn}}, \ and\
  \bibinfo {author} {\bibfnamefont {P.-G.}\ \bibnamefont {Reinhard}},\ }\href
  {http://link.aps.org/doi/10.1103/PhysRevC.69.014307} {\bibfield  {journal}
  {\bibinfo  {journal} {Phys. Rev. C}\ }\textbf {\bibinfo {volume} {69}},\
  \bibinfo {pages} {014307} (\bibinfo {year} {2004})}\BibitemShut {NoStop}%
\bibitem [{\citenamefont {Blum}\ \emph {et~al.}(1994)\citenamefont {Blum},
  \citenamefont {Maruhn}, \citenamefont {Reinhard},\ and\ \citenamefont
  {Greiner}}]{Blum1994_PLB323-262}%
  \BibitemOpen
  \bibfield  {author} {\bibinfo {author} {\bibfnamefont {V.}~\bibnamefont
  {Blum}}, \bibinfo {author} {\bibfnamefont {J.}~\bibnamefont {Maruhn}},
  \bibinfo {author} {\bibfnamefont {P.-G.}\ \bibnamefont {Reinhard}}, \ and\
  \bibinfo {author} {\bibfnamefont {W.}~\bibnamefont {Greiner}},\ }\href
  {http://www.sciencedirect.com/science/article/pii/0370269394912173}
  {\bibfield  {journal} {\bibinfo  {journal} {Phys. Lett. B}\ }\textbf
  {\bibinfo {volume} {323}},\ \bibinfo {pages} {262} (\bibinfo {year}
  {1994})}\BibitemShut {NoStop}%
\bibitem [{\citenamefont {Zhang}\ \emph {et~al.}(2003)\citenamefont {Zhang},
  \citenamefont {Zhang}, \citenamefont {Zhang},\ and\ \citenamefont
  {Meng}}]{Zhang2003_CPL20-1694}%
  \BibitemOpen
  \bibfield  {author} {\bibinfo {author} {\bibfnamefont {W.}~\bibnamefont
  {Zhang}}, \bibinfo {author} {\bibfnamefont {S.-S.}\ \bibnamefont {Zhang}},
  \bibinfo {author} {\bibfnamefont {S.-Q.}\ \bibnamefont {Zhang}}, \ and\
  \bibinfo {author} {\bibfnamefont {J.}~\bibnamefont {Meng}},\ }\href
  {http://stacks.iop.org/0256-307X/20/i=10/a=312} {\bibfield  {journal}
  {\bibinfo  {journal} {Chin. Phys. Lett.}\ }\textbf {\bibinfo {volume} {20}},\
  \bibinfo {pages} {1694} (\bibinfo {year} {2003})}\BibitemShut {NoStop}%
\bibitem [{\citenamefont {Bender}\ \emph {et~al.}(2003)\citenamefont {Bender},
  \citenamefont {Heenen},\ and\ \citenamefont
  {Reinhard}}]{Bender2003_RMP75-121}%
  \BibitemOpen
  \bibfield  {author} {\bibinfo {author} {\bibfnamefont {M.}~\bibnamefont
  {Bender}}, \bibinfo {author} {\bibfnamefont {P.-H.}\ \bibnamefont {Heenen}},
  \ and\ \bibinfo {author} {\bibfnamefont {P.-G.}\ \bibnamefont {Reinhard}},\
  }\href {http://link.aps.org/doi/10.1103/RevModPhys.75.121} {\bibfield
  {journal} {\bibinfo  {journal} {Rev. Mod. Phys.}\ }\textbf {\bibinfo {volume}
  {75}},\ \bibinfo {pages} {121} (\bibinfo {year} {2003})}\BibitemShut
  {NoStop}%
\bibitem [{\citenamefont {Lu}\ \emph {et~al.}(2006)\citenamefont {Lu},
  \citenamefont {Geng},\ and\ \citenamefont {Meng}}]{Lu2006_CPL23-2940}%
  \BibitemOpen
  \bibfield  {author} {\bibinfo {author} {\bibfnamefont {H.-F.}\ \bibnamefont
  {Lu}}, \bibinfo {author} {\bibfnamefont {L.-S.}\ \bibnamefont {Geng}}, \ and\
  \bibinfo {author} {\bibfnamefont {J.}~\bibnamefont {Meng}},\ }\href
  {http://stacks.iop.org/0256-307X/23/i=11/a=016} {\bibfield  {journal}
  {\bibinfo  {journal} {Chin. Phys. Lett.}\ }\textbf {\bibinfo {volume} {23}},\
  \bibinfo {pages} {2940} (\bibinfo {year} {2006})}\BibitemShut {NoStop}%
\bibitem [{\citenamefont {Li}\ \emph {et~al.}(2010)\citenamefont {Li},
  \citenamefont {Nik\v{s}i\'c}, \citenamefont {Vretenar}, \citenamefont
  {Ring},\ and\ \citenamefont {Meng}}]{Li2010_PRC81-064321}%
  \BibitemOpen
  \bibfield  {author} {\bibinfo {author} {\bibfnamefont {Z.~P.}\ \bibnamefont
  {Li}}, \bibinfo {author} {\bibfnamefont {T.}~\bibnamefont {Nik\v{s}i\'c}},
  \bibinfo {author} {\bibfnamefont {D.}~\bibnamefont {Vretenar}}, \bibinfo
  {author} {\bibfnamefont {P.}~\bibnamefont {Ring}}, \ and\ \bibinfo {author}
  {\bibfnamefont {J.}~\bibnamefont {Meng}},\ }\href
  {http://link.aps.org/doi/10.1103/PhysRevC.81.064321} {\bibfield  {journal}
  {\bibinfo  {journal} {Phys. Rev. C}\ }\textbf {\bibinfo {volume} {81}},\
  \bibinfo {pages} {064321} (\bibinfo {year} {2010})}\BibitemShut {NoStop}%
\bibitem [{\citenamefont {Abusara}\ \emph {et~al.}(2010)\citenamefont
  {Abusara}, \citenamefont {Afanasjev},\ and\ \citenamefont
  {Ring}}]{Abusara2010_PRC82-044303}%
  \BibitemOpen
  \bibfield  {author} {\bibinfo {author} {\bibfnamefont {H.}~\bibnamefont
  {Abusara}}, \bibinfo {author} {\bibfnamefont {A.~V.}\ \bibnamefont
  {Afanasjev}}, \ and\ \bibinfo {author} {\bibfnamefont {P.}~\bibnamefont
  {Ring}},\ }\href {http://link.aps.org/doi/10.1103/PhysRevC.82.044303}
  {\bibfield  {journal} {\bibinfo  {journal} {Phys. Rev. C}\ }\textbf {\bibinfo
  {volume} {82}},\ \bibinfo {pages} {044303} (\bibinfo {year}
  {2010})}\BibitemShut {NoStop}%
\bibitem [{\citenamefont {Abusara}\ \emph {et~al.}(2012)\citenamefont
  {Abusara}, \citenamefont {Afanasjev},\ and\ \citenamefont
  {Ring}}]{Abusara2012_PRC85-024314}%
  \BibitemOpen
  \bibfield  {author} {\bibinfo {author} {\bibfnamefont {H.}~\bibnamefont
  {Abusara}}, \bibinfo {author} {\bibfnamefont {A.~V.}\ \bibnamefont
  {Afanasjev}}, \ and\ \bibinfo {author} {\bibfnamefont {P.}~\bibnamefont
  {Ring}},\ }\href {http://link.aps.org/doi/10.1103/PhysRevC.85.024314}
  {\bibfield  {journal} {\bibinfo  {journal} {Phys. Rev. C}\ }\textbf {\bibinfo
  {volume} {85}},\ \bibinfo {pages} {024314} (\bibinfo {year}
  {2012})}\BibitemShut {NoStop}%
\bibitem [{\citenamefont {Lu}\ \emph {et~al.}(2012)\citenamefont {Lu},
  \citenamefont {Zhao},\ and\ \citenamefont {Zhou}}]{Lu2012_PRC85-011301R}%
  \BibitemOpen
  \bibfield  {author} {\bibinfo {author} {\bibfnamefont {B.-N.}\ \bibnamefont
  {Lu}}, \bibinfo {author} {\bibfnamefont {E.-G.}\ \bibnamefont {Zhao}}, \ and\
  \bibinfo {author} {\bibfnamefont {S.-G.}\ \bibnamefont {Zhou}},\ }\href
  {http://link.aps.org/doi/10.1103/PhysRevC.85.011301} {\bibfield  {journal}
  {\bibinfo  {journal} {Phys. Rev. C}\ }\textbf {\bibinfo {volume} {85}},\
  \bibinfo {pages} {011301(R)} (\bibinfo {year} {2012})}\BibitemShut {NoStop}%
\bibitem [{\citenamefont {Lu}\ \emph {et~al.}(2014)\citenamefont {Lu},
  \citenamefont {Zhao}, \citenamefont {Zhao},\ and\ \citenamefont
  {Zhou}}]{Lu2014_PRC89-014323}%
  \BibitemOpen
  \bibfield  {author} {\bibinfo {author} {\bibfnamefont {B.-N.}\ \bibnamefont
  {Lu}}, \bibinfo {author} {\bibfnamefont {J.}~\bibnamefont {Zhao}}, \bibinfo
  {author} {\bibfnamefont {E.-G.}\ \bibnamefont {Zhao}}, \ and\ \bibinfo
  {author} {\bibfnamefont {S.-G.}\ \bibnamefont {Zhou}},\ }\href
  {http://link.aps.org/doi/10.1103/PhysRevC.89.014323} {\bibfield  {journal}
  {\bibinfo  {journal} {Phys. Rev. C}\ }\textbf {\bibinfo {volume} {89}},\
  \bibinfo {pages} {014323} (\bibinfo {year} {2014})}\BibitemShut {NoStop}%
\bibitem [{\citenamefont {Zhao}\ \emph
  {et~al.}(2015{\natexlab{a}})\citenamefont {Zhao}, \citenamefont {Lu},
  \citenamefont {Vretenar}, \citenamefont {Zhao},\ and\ \citenamefont
  {Zhou}}]{Zhao2015_PRC91-014321}%
  \BibitemOpen
  \bibfield  {author} {\bibinfo {author} {\bibfnamefont {J.}~\bibnamefont
  {Zhao}}, \bibinfo {author} {\bibfnamefont {B.-N.}\ \bibnamefont {Lu}},
  \bibinfo {author} {\bibfnamefont {D.}~\bibnamefont {Vretenar}}, \bibinfo
  {author} {\bibfnamefont {E.-G.}\ \bibnamefont {Zhao}}, \ and\ \bibinfo
  {author} {\bibfnamefont {S.-G.}\ \bibnamefont {Zhou}},\ }\href
  {http://link.aps.org/doi/10.1103/PhysRevC.91.014321} {\bibfield  {journal}
  {\bibinfo  {journal} {Phys. Rev. C}\ }\textbf {\bibinfo {volume} {91}},\
  \bibinfo {pages} {014321} (\bibinfo {year} {2015}{\natexlab{a}})}\BibitemShut
  {NoStop}%
\bibitem [{\citenamefont {Agbemava}\ \emph {et~al.}(2017)\citenamefont
  {Agbemava}, \citenamefont {Afanasjev}, \citenamefont {Ray},\ and\
  \citenamefont {Ring}}]{Agbemava2017_PRC95-054324}%
  \BibitemOpen
  \bibfield  {author} {\bibinfo {author} {\bibfnamefont {S.~E.}\ \bibnamefont
  {Agbemava}}, \bibinfo {author} {\bibfnamefont {A.~V.}\ \bibnamefont
  {Afanasjev}}, \bibinfo {author} {\bibfnamefont {D.}~\bibnamefont {Ray}}, \
  and\ \bibinfo {author} {\bibfnamefont {P.}~\bibnamefont {Ring}},\ }\href
  {https://link.aps.org/doi/10.1103/PhysRevC.95.054324} {\bibfield  {journal}
  {\bibinfo  {journal} {Phys. Rev. C}\ }\textbf {\bibinfo {volume} {95}},\
  \bibinfo {pages} {054324} (\bibinfo {year} {2017})}\BibitemShut {NoStop}%
\bibitem [{\citenamefont {Prassa}\ \emph {et~al.}(2012)\citenamefont {Prassa},
  \citenamefont {Nik\v{s}i\'c}, \citenamefont {Lalazissis},\ and\ \citenamefont
  {Vretenar}}]{Prassa2012_PRC86-024317}%
  \BibitemOpen
  \bibfield  {author} {\bibinfo {author} {\bibfnamefont {V.}~\bibnamefont
  {Prassa}}, \bibinfo {author} {\bibfnamefont {T.}~\bibnamefont
  {Nik\v{s}i\'c}}, \bibinfo {author} {\bibfnamefont {G.~A.}\ \bibnamefont
  {Lalazissis}}, \ and\ \bibinfo {author} {\bibfnamefont {D.}~\bibnamefont
  {Vretenar}},\ }\href {http://link.aps.org/doi/10.1103/PhysRevC.86.024317}
  {\bibfield  {journal} {\bibinfo  {journal} {Phys. Rev. C}\ }\textbf {\bibinfo
  {volume} {86}},\ \bibinfo {pages} {024317} (\bibinfo {year}
  {2012})}\BibitemShut {NoStop}%
\bibitem [{\citenamefont {Zhao}\ \emph
  {et~al.}(2015{\natexlab{b}})\citenamefont {Zhao}, \citenamefont {Lu},
  \citenamefont {Niksic},\ and\ \citenamefont
  {Vretenar}}]{Zhao2015_PRC92-064315}%
  \BibitemOpen
  \bibfield  {author} {\bibinfo {author} {\bibfnamefont {J.}~\bibnamefont
  {Zhao}}, \bibinfo {author} {\bibfnamefont {B.-N.}\ \bibnamefont {Lu}},
  \bibinfo {author} {\bibfnamefont {T.}~\bibnamefont {Niksic}}, \ and\ \bibinfo
  {author} {\bibfnamefont {D.}~\bibnamefont {Vretenar}},\ }\href
  {http://link.aps.org/doi/10.1103/PhysRevC.92.064315} {\bibfield  {journal}
  {\bibinfo  {journal} {Phys. Rev. C}\ }\textbf {\bibinfo {volume} {92}},\
  \bibinfo {pages} {064315} (\bibinfo {year} {2015}{\natexlab{b}})}\BibitemShut
  {NoStop}%
\bibitem [{\citenamefont {Zhao}\ \emph {et~al.}(2016)\citenamefont {Zhao},
  \citenamefont {Lu}, \citenamefont {Niksic}, \citenamefont {Vretenar},\ and\
  \citenamefont {Zhou}}]{Zhao2016_PRC93-044315}%
  \BibitemOpen
  \bibfield  {author} {\bibinfo {author} {\bibfnamefont {J.}~\bibnamefont
  {Zhao}}, \bibinfo {author} {\bibfnamefont {B.-N.}\ \bibnamefont {Lu}},
  \bibinfo {author} {\bibfnamefont {T.}~\bibnamefont {Niksic}}, \bibinfo
  {author} {\bibfnamefont {D.}~\bibnamefont {Vretenar}}, \ and\ \bibinfo
  {author} {\bibfnamefont {S.-G.}\ \bibnamefont {Zhou}},\ }\href
  {http://link.aps.org/doi/10.1103/PhysRevC.93.044315} {\bibfield  {journal}
  {\bibinfo  {journal} {Phys. Rev. C}\ }\textbf {\bibinfo {volume} {93}},\
  \bibinfo {pages} {044315} (\bibinfo {year} {2016})}\BibitemShut {NoStop}%
\bibitem [{\citenamefont {Zhao}\ \emph {et~al.}(2017)\citenamefont {Zhao},
  \citenamefont {Lu}, \citenamefont {Zhao},\ and\ \citenamefont
  {Zhou}}]{Zhao2017_PRC95-014320}%
  \BibitemOpen
  \bibfield  {author} {\bibinfo {author} {\bibfnamefont {J.}~\bibnamefont
  {Zhao}}, \bibinfo {author} {\bibfnamefont {B.-N.}\ \bibnamefont {Lu}},
  \bibinfo {author} {\bibfnamefont {E.-G.}\ \bibnamefont {Zhao}}, \ and\
  \bibinfo {author} {\bibfnamefont {S.-G.}\ \bibnamefont {Zhou}},\ }\href
  {\doibase 10.1103/PhysRevC.95.014320} {\bibfield  {journal} {\bibinfo
  {journal} {Phys. Rev. C}\ }\textbf {\bibinfo {volume} {95}},\ \bibinfo
  {pages} {014320} (\bibinfo {year} {2017})}\BibitemShut {NoStop}%
\bibitem [{\citenamefont {Tao}\ \emph {et~al.}(2017)\citenamefont {Tao},
  \citenamefont {Zhao}, \citenamefont {Li}, \citenamefont {Niksic},\ and\
  \citenamefont {Vretenar}}]{Tao2017_PRC96-024319}%
  \BibitemOpen
  \bibfield  {author} {\bibinfo {author} {\bibfnamefont {H.}~\bibnamefont
  {Tao}}, \bibinfo {author} {\bibfnamefont {J.}~\bibnamefont {Zhao}}, \bibinfo
  {author} {\bibfnamefont {Z.~P.}\ \bibnamefont {Li}}, \bibinfo {author}
  {\bibfnamefont {T.}~\bibnamefont {Niksic}}, \ and\ \bibinfo {author}
  {\bibfnamefont {D.}~\bibnamefont {Vretenar}},\ }\href
  {https://link.aps.org/doi/10.1103/PhysRevC.96.024319} {\bibfield  {journal}
  {\bibinfo  {journal} {Phys. Rev. C}\ }\textbf {\bibinfo {volume} {96}},\
  \bibinfo {pages} {024319} (\bibinfo {year} {2017})}\BibitemShut {NoStop}%
\bibitem [{\citenamefont {Goodman}(1981)}]{Goodman1981_NPA352-30}%
  \BibitemOpen
  \bibfield  {author} {\bibinfo {author} {\bibfnamefont {A.~L.}\ \bibnamefont
  {Goodman}},\ }\href
  {http://www.sciencedirect.com/science/article/pii/0375947481905571}
  {\bibfield  {journal} {\bibinfo  {journal} {Nucl. Phys. A}\ }\textbf
  {\bibinfo {volume} {352}},\ \bibinfo {pages} {30} (\bibinfo {year}
  {1981})}\BibitemShut {NoStop}%
\bibitem [{\citenamefont {Zhu}\ and\ \citenamefont
  {Pei}(2016)}]{Zhu2016_PRC94-024329}%
  \BibitemOpen
  \bibfield  {author} {\bibinfo {author} {\bibfnamefont {Y.}~\bibnamefont
  {Zhu}}\ and\ \bibinfo {author} {\bibfnamefont {J.~C.}\ \bibnamefont {Pei}},\
  }\href {http://link.aps.org/doi/10.1103/PhysRevC.94.024329} {\bibfield
  {journal} {\bibinfo  {journal} {Phys. Rev. C}\ }\textbf {\bibinfo {volume}
  {94}},\ \bibinfo {pages} {024329} (\bibinfo {year} {2016})}\BibitemShut
  {NoStop}%
\bibitem [{\citenamefont {Schunck}\ \emph {et~al.}(2015)\citenamefont
  {Schunck}, \citenamefont {Duke},\ and\ \citenamefont
  {Carr}}]{Schunck2015_PRC91-034327}%
  \BibitemOpen
  \bibfield  {author} {\bibinfo {author} {\bibfnamefont {N.}~\bibnamefont
  {Schunck}}, \bibinfo {author} {\bibfnamefont {D.}~\bibnamefont {Duke}}, \
  and\ \bibinfo {author} {\bibfnamefont {H.}~\bibnamefont {Carr}},\ }\href
  {http://link.aps.org/doi/10.1103/PhysRevC.91.034327} {\bibfield  {journal}
  {\bibinfo  {journal} {Phys. Rev. C}\ }\textbf {\bibinfo {volume} {91}},\
  \bibinfo {pages} {034327} (\bibinfo {year} {2015})}\BibitemShut {NoStop}%
\bibitem [{\citenamefont {McDonnell}\ \emph {et~al.}(2013)\citenamefont
  {McDonnell}, \citenamefont {Nazarewicz},\ and\ \citenamefont
  {Sheikh}}]{McDonnell2013_PRC87-054327}%
  \BibitemOpen
  \bibfield  {author} {\bibinfo {author} {\bibfnamefont {J.~D.}\ \bibnamefont
  {McDonnell}}, \bibinfo {author} {\bibfnamefont {W.}~\bibnamefont
  {Nazarewicz}}, \ and\ \bibinfo {author} {\bibfnamefont {J.~A.}\ \bibnamefont
  {Sheikh}},\ }\href {http://link.aps.org/doi/10.1103/PhysRevC.87.054327}
  {\bibfield  {journal} {\bibinfo  {journal} {Phys. Rev. C}\ }\textbf {\bibinfo
  {volume} {87}},\ \bibinfo {pages} {054327} (\bibinfo {year}
  {2013})}\BibitemShut {NoStop}%
\bibitem [{\citenamefont {McDonnell}\ \emph {et~al.}(2014)\citenamefont
  {McDonnell}, \citenamefont {Nazarewicz}, \citenamefont {Sheikh},
  \citenamefont {Staszczak},\ and\ \citenamefont
  {Warda}}]{McDonnell2014_PRC90-021302R}%
  \BibitemOpen
  \bibfield  {author} {\bibinfo {author} {\bibfnamefont {J.~D.}\ \bibnamefont
  {McDonnell}}, \bibinfo {author} {\bibfnamefont {W.}~\bibnamefont
  {Nazarewicz}}, \bibinfo {author} {\bibfnamefont {J.~A.}\ \bibnamefont
  {Sheikh}}, \bibinfo {author} {\bibfnamefont {A.}~\bibnamefont {Staszczak}}, \
  and\ \bibinfo {author} {\bibfnamefont {M.}~\bibnamefont {Warda}},\ }\href
  {http://link.aps.org/doi/10.1103/PhysRevC.90.021302} {\bibfield  {journal}
  {\bibinfo  {journal} {Phys. Rev. C}\ }\textbf {\bibinfo {volume} {90}},\
  \bibinfo {pages} {021302(R)} (\bibinfo {year} {2014})}\BibitemShut {NoStop}%
\bibitem [{\citenamefont {Pei}\ \emph {et~al.}(2009)\citenamefont {Pei},
  \citenamefont {Nazarewicz}, \citenamefont {Sheikh},\ and\ \citenamefont
  {Kerman}}]{Pei2009_PRL102-192501}%
  \BibitemOpen
  \bibfield  {author} {\bibinfo {author} {\bibfnamefont {J.~C.}\ \bibnamefont
  {Pei}}, \bibinfo {author} {\bibfnamefont {W.}~\bibnamefont {Nazarewicz}},
  \bibinfo {author} {\bibfnamefont {J.~A.}\ \bibnamefont {Sheikh}}, \ and\
  \bibinfo {author} {\bibfnamefont {A.~K.}\ \bibnamefont {Kerman}},\ }\href
  {http://link.aps.org/doi/10.1103/PhysRevLett.102.192501} {\bibfield
  {journal} {\bibinfo  {journal} {Phys. Rev. Lett.}\ }\textbf {\bibinfo
  {volume} {102}},\ \bibinfo {pages} {192501} (\bibinfo {year}
  {2009})}\BibitemShut {NoStop}%
\bibitem [{\citenamefont {Martin}\ and\ \citenamefont
  {Robledo}(2009)}]{Martin2009_IJMPE18-861}%
  \BibitemOpen
  \bibfield  {author} {\bibinfo {author} {\bibfnamefont {V.}~\bibnamefont
  {Martin}}\ and\ \bibinfo {author} {\bibfnamefont {L.~M.}\ \bibnamefont
  {Robledo}},\ }\href {\doibase 10.1142/s0218301309012963} {\bibfield
  {journal} {\bibinfo  {journal} {Int. J. Mod. Phys. E}\ }\textbf {\bibinfo
  {volume} {18}},\ \bibinfo {pages} {861} (\bibinfo {year} {2009})}\BibitemShut
  {NoStop}%
\bibitem [{\citenamefont {Ivanyuk}\ \emph {et~al.}(2018)\citenamefont
  {Ivanyuk}, \citenamefont {Ishizuka}, \citenamefont {Usang},\ and\
  \citenamefont {Chiba}}]{Ivanyuk2018_PRC97-054331}%
  \BibitemOpen
  \bibfield  {author} {\bibinfo {author} {\bibfnamefont {F.~A.}\ \bibnamefont
  {Ivanyuk}}, \bibinfo {author} {\bibfnamefont {C.}~\bibnamefont {Ishizuka}},
  \bibinfo {author} {\bibfnamefont {M.~D.}\ \bibnamefont {Usang}}, \ and\
  \bibinfo {author} {\bibfnamefont {S.}~\bibnamefont {Chiba}},\ }\href
  {https://link.aps.org/doi/10.1103/PhysRevC.97.054331} {\bibfield  {journal}
  {\bibinfo  {journal} {Phys. Rev. C}\ }\textbf {\bibinfo {volume} {97}},\
  \bibinfo {pages} {054331} (\bibinfo {year} {2018})}\BibitemShut {NoStop}%
\bibitem [{\citenamefont {Randrup}\ and\ \citenamefont
  {M\"oller}(2013)}]{Randrup2013_PRC88-064606}%
  \BibitemOpen
  \bibfield  {author} {\bibinfo {author} {\bibfnamefont {J.}~\bibnamefont
  {Randrup}}\ and\ \bibinfo {author} {\bibfnamefont {P.}~\bibnamefont
  {M\"oller}},\ }\href {http://link.aps.org/doi/10.1103/PhysRevC.88.064606}
  {\bibfield  {journal} {\bibinfo  {journal} {Phys. Rev. C}\ }\textbf {\bibinfo
  {volume} {88}},\ \bibinfo {pages} {064606} (\bibinfo {year}
  {2013})}\BibitemShut {NoStop}%
\bibitem [{\citenamefont {Pasca}\ \emph {et~al.}(2016)\citenamefont {Pasca},
  \citenamefont {Andreev}, \citenamefont {Adamian},\ and\ \citenamefont
  {Antonenko}}]{Pasca2016_PLB760-800}%
  \BibitemOpen
  \bibfield  {author} {\bibinfo {author} {\bibfnamefont {H.}~\bibnamefont
  {Pasca}}, \bibinfo {author} {\bibfnamefont {A.~V.}\ \bibnamefont {Andreev}},
  \bibinfo {author} {\bibfnamefont {G.~G.}\ \bibnamefont {Adamian}}, \ and\
  \bibinfo {author} {\bibfnamefont {N.~V.}\ \bibnamefont {Antonenko}},\ }\href
  {http://www.sciencedirect.com/science/article/pii/S0370269316304154}
  {\bibfield  {journal} {\bibinfo  {journal} {Phys. Lett. B}\ }\textbf
  {\bibinfo {volume} {760}},\ \bibinfo {pages} {800} (\bibinfo {year}
  {2016})}\BibitemShut {NoStop}%
\bibitem [{\citenamefont {Egido}\ \emph {et~al.}(1986)\citenamefont {Egido},
  \citenamefont {Ring},\ and\ \citenamefont {Mang}}]{Egido1986_NPA451-77}%
  \BibitemOpen
  \bibfield  {author} {\bibinfo {author} {\bibfnamefont {J.~L.}\ \bibnamefont
  {Egido}}, \bibinfo {author} {\bibfnamefont {P.}~\bibnamefont {Ring}}, \ and\
  \bibinfo {author} {\bibfnamefont {H.~J.}\ \bibnamefont {Mang}},\ }\href
  {http://www.sciencedirect.com/science/article/pii/0375947486902423}
  {\bibfield  {journal} {\bibinfo  {journal} {Nucl. Phys. A}\ }\textbf
  {\bibinfo {volume} {451}},\ \bibinfo {pages} {77} (\bibinfo {year}
  {1986})}\BibitemShut {NoStop}%
\bibitem [{\citenamefont {Nik\v{s}i\'c}\ \emph {et~al.}(2008)\citenamefont
  {Nik\v{s}i\'c}, \citenamefont {Vretenar},\ and\ \citenamefont
  {Ring}}]{Niksic2008_PRC78-034318}%
  \BibitemOpen
  \bibfield  {author} {\bibinfo {author} {\bibfnamefont {T.}~\bibnamefont
  {Nik\v{s}i\'c}}, \bibinfo {author} {\bibfnamefont {D.}~\bibnamefont
  {Vretenar}}, \ and\ \bibinfo {author} {\bibfnamefont {P.}~\bibnamefont
  {Ring}},\ }\href {http://link.aps.org/doi/10.1103/PhysRevC.78.034318}
  {\bibfield  {journal} {\bibinfo  {journal} {Phys. Rev. C}\ }\textbf {\bibinfo
  {volume} {78}},\ \bibinfo {pages} {034318} (\bibinfo {year}
  {2008})}\BibitemShut {NoStop}%
\bibitem [{\citenamefont {Tian}\ \emph {et~al.}(2009)\citenamefont {Tian},
  \citenamefont {Ma},\ and\ \citenamefont {Ring}}]{Tian2009_PLB676-44}%
  \BibitemOpen
  \bibfield  {author} {\bibinfo {author} {\bibfnamefont {Y.}~\bibnamefont
  {Tian}}, \bibinfo {author} {\bibfnamefont {Z.~Y.}\ \bibnamefont {Ma}}, \ and\
  \bibinfo {author} {\bibfnamefont {P.}~\bibnamefont {Ring}},\ }\href
  {http://www.sciencedirect.com/science/article/pii/S0370269309004912}
  {\bibfield  {journal} {\bibinfo  {journal} {Phys. Lett. B}\ }\textbf
  {\bibinfo {volume} {676}},\ \bibinfo {pages} {44} (\bibinfo {year}
  {2009})}\BibitemShut {NoStop}%
\bibitem [{\citenamefont {Gambhir}\ \emph {et~al.}(1990)\citenamefont
  {Gambhir}, \citenamefont {Ring},\ and\ \citenamefont
  {Thimet}}]{Gambhir1990_APNY198-132}%
  \BibitemOpen
  \bibfield  {author} {\bibinfo {author} {\bibfnamefont {Y.}~\bibnamefont
  {Gambhir}}, \bibinfo {author} {\bibfnamefont {P.}~\bibnamefont {Ring}}, \
  and\ \bibinfo {author} {\bibfnamefont {A.}~\bibnamefont {Thimet}},\ }\href
  {http://www.sciencedirect.com/science/article/pii/000349169090330Q}
  {\bibfield  {journal} {\bibinfo  {journal} {Ann. Phys.}\ }\textbf {\bibinfo
  {volume} {198}},\ \bibinfo {pages} {132} (\bibinfo {year}
  {1990})}\BibitemShut {NoStop}%
\bibitem [{\citenamefont {Wang}\ \emph {et~al.}(2013)\citenamefont {Wang},
  \citenamefont {Sun}, \citenamefont {Dong},\ and\ \citenamefont
  {Long}}]{Wang2013_PRC87-054331}%
  \BibitemOpen
  \bibfield  {author} {\bibinfo {author} {\bibfnamefont {L.~J.}\ \bibnamefont
  {Wang}}, \bibinfo {author} {\bibfnamefont {B.~Y.}\ \bibnamefont {Sun}},
  \bibinfo {author} {\bibfnamefont {J.~M.}\ \bibnamefont {Dong}}, \ and\
  \bibinfo {author} {\bibfnamefont {W.~H.}\ \bibnamefont {Long}},\ }\href
  {http://link.aps.org/doi/10.1103/PhysRevC.87.054331} {\bibfield  {journal}
  {\bibinfo  {journal} {Phys. Rev. C}\ }\textbf {\bibinfo {volume} {87}},\
  \bibinfo {pages} {054331} (\bibinfo {year} {2013})}\BibitemShut {NoStop}%
\bibitem [{\citenamefont {Afanasjev}\ and\ \citenamefont
  {Abdurazakov}(2013)}]{Afanasjev2013_PRC88-014320}%
  \BibitemOpen
  \bibfield  {author} {\bibinfo {author} {\bibfnamefont {A.~V.}\ \bibnamefont
  {Afanasjev}}\ and\ \bibinfo {author} {\bibfnamefont {O.}~\bibnamefont
  {Abdurazakov}},\ }\href {http://link.aps.org/doi/10.1103/PhysRevC.88.014320}
  {\bibfield  {journal} {\bibinfo  {journal} {Phys. Rev. C}\ }\textbf {\bibinfo
  {volume} {88}},\ \bibinfo {pages} {014320} (\bibinfo {year}
  {2013})}\BibitemShut {NoStop}%
\bibitem [{\citenamefont {Younes}\ and\ \citenamefont
  {Gogny}(2009)}]{Younes2009_PRC80-054313}%
  \BibitemOpen
  \bibfield  {author} {\bibinfo {author} {\bibfnamefont {W.}~\bibnamefont
  {Younes}}\ and\ \bibinfo {author} {\bibfnamefont {D.}~\bibnamefont {Gogny}},\
  }\href {http://link.aps.org/doi/10.1103/PhysRevC.80.054313} {\bibfield
  {journal} {\bibinfo  {journal} {Phys. Rev. C}\ }\textbf {\bibinfo {volume}
  {80}},\ \bibinfo {pages} {054313} (\bibinfo {year} {2009})}\BibitemShut
  {NoStop}%
\bibitem [{\citenamefont {Zhao}\ \emph {et~al.}(2010)\citenamefont {Zhao},
  \citenamefont {Li}, \citenamefont {Yao},\ and\ \citenamefont
  {Meng}}]{Zhao2010_PRC82-054319}%
  \BibitemOpen
  \bibfield  {author} {\bibinfo {author} {\bibfnamefont {P.~W.}\ \bibnamefont
  {Zhao}}, \bibinfo {author} {\bibfnamefont {Z.~P.}\ \bibnamefont {Li}},
  \bibinfo {author} {\bibfnamefont {J.~M.}\ \bibnamefont {Yao}}, \ and\
  \bibinfo {author} {\bibfnamefont {J.}~\bibnamefont {Meng}},\ }\href
  {http://link.aps.org/doi/10.1103/PhysRevC.82.054319} {\bibfield  {journal}
  {\bibinfo  {journal} {Phys. Rev. C}\ }\textbf {\bibinfo {volume} {82}},\
  \bibinfo {pages} {054319} (\bibinfo {year} {2010})}\BibitemShut {NoStop}%
\bibitem [{\citenamefont {Moretto}\ and\ \citenamefont
  {Babinet}(1974)}]{Moretto1974_PLB49-147}%
  \BibitemOpen
  \bibfield  {author} {\bibinfo {author} {\bibfnamefont {L.}~\bibnamefont
  {Moretto}}\ and\ \bibinfo {author} {\bibfnamefont {R.}~\bibnamefont
  {Babinet}},\ }\href
  {http://www.sciencedirect.com/science/article/pii/0370269374904948}
  {\bibfield  {journal} {\bibinfo  {journal} {Phys. Lett. B}\ }\textbf
  {\bibinfo {volume} {49}},\ \bibinfo {pages} {147} (\bibinfo {year}
  {1974})}\BibitemShut {NoStop}%
\end{thebibliography}


%

\end{document}